# In silicio stretching of chromatin

**Frank Aumann[1], Filip Lankas[2], Maïwen Caudron[3] and Jörg Langowski[1*]**

[1] ***Division Biophysics of Macromolecules, German Cancer Research Center,***

***Im Neuenheimer Feld 580, D-69120 Heidelberg, Germany***

[2] ***Swiss Federal Institute of Technology,***

***Station 8, CH-1015 Lausanne, Switzerland***

[3] ***European Molecular Biology Laboratory,***

***Meyerhofstrasse 1, Postfach 102209, D-69117 Heidelberg, Germany***

* *Corresponding author:* Tel.: +49 6221 42 3393; Fax: +49 6221 4252 3390; email: jl@dkfz.de

Running title: *Chromatin elasticity*

Abbreviations used: NCP, nucleosome core particle, $L_p$, persistence length, MC, Monte Carlo.

# Abstract

We present Monte-Carlo (MC) simulations of the stretching of a single 30 nm chromatin fiber. The model approximates the DNA by a flexible polymer chain with Debye-Hückel electrostatics and uses a two-angle zig-zag model for the geometry of the linker DNA connecting the nucleosomes. The latter are represented by flat disks interacting via an attractive Gay-Berne potential. Our results show that the stiffness of the chromatin fiber strongly depends on the linker DNA length. Furthermore, changing the twisting angle between nucleosomes from 90° to 130° increases the stiffness significantly. An increase in the opening angle from 22° to 34° leads to softer fibers for small linker lengths. We observe that fibers containing a linker histone at each nucleosome are stiffer compared to those without the linker histone. The simulated persistence lengths and elastic moduli agree with experimental data. Finally, we show that the chromatin fiber does not behave as an isotropic elastic rod, but its rigidity depends on the direction of deformation: chromatin is much more resistant to stretching than to bending.

# Introduction

DNA in eukaryotic cells is complexed with histone proteins into chromatin. Apart from the compaction of DNA into chromatin itself, changes in the chromatin structure play a major role in the regulation of gene expression [1, 2], nuclear architecture [3, 4] or DNA-protein interactions in general [5]. In order to understand the molecular basis for the structural variability of chromatin, the mechanical properties of the chromatin fiber have been studied recently in a number of single molecule experiments [6-10]. Computer simulations of the chromatin fiber can help us to interpret this experimental data and estimate the bending and stretching stiffness of the chromatin fiber. The spatial distribution and the dynamics of genetic material in the interphase nucleus strongly depend on these nanomechanical parameters.

In most eukaryotic organisms the first level of genome compaction in the nucleus is the packaging of DNA into the nucleosome core particle (NCP). X-ray crystallography [11] revealed the shape of this complex as a flat disk, with a diameter of 11 nm and a height of 5.5 nm. It consists of 146 base pairs (bp) of DNA wrapped 1.65 times around a histone protein octamer formed by dimers of histone H2A, H2B, H3 and H4. The basic unit of chromatin, the nucleosome, consists of one NCP and the stretch of DNA linking it to the next one [12]. Furthermore, the linker histone H1 may be attached to the NCP, forming a stem-motif. This complex is called a chromatosome.

The “bead-chain” structure of nucleosomes [13-15] spaced regularly on DNA at low salt concentrations, referred to as the “10 nm fiber” [16] , will compact under physiological conditions into a fiber-like structure with a diameter of approximately 30 nm [17], the so called “30 nm fiber” [18]. Although the structure of the nucleosome is known from X-ray crystallography [11] up to an atomic resolution of 1.9 Å [19], the precise arrangement of DNA and histones inside the 30 nm fiber is still controversial. The current discussion focuses on two classes of models: the s*olenoid models* [16, 18, 20, 21] and the *zig-zag models* [22-27].

The solenoid model assumes a helical structure for the chain of nucleosomes with their flat faces roughly parallel to the long axis of the fiber. The DNA entry-exit side points inward to the axis of the solenoid. To connect two neighboring nucleosomes, the linker DNA has to be bent and therefore substantial input of elastic energy is required. This energy input is not necessary in zig-zag-models. Here, straight linker DNA is assumed to connect two nucleosomes located at the opposite sides of the fiber. In a three-dimensional view the DNA follows a zig-zag-like path.

Cyro electron microscopy [11, 26, 28] observed a zig-zag motif at lower salt concentration which became more compact when approaching a physiological value. These compacted structures look similar to the conformations proposed by the zig-zag model of Woodcock et al. [22]. Mass density, linker entry-exit angles and other structural parameters obtained by atomic force microscopy [25, 29], neutron scattering and scanning transmission electron microscopy [30] also confirm the zig-zag models. Very recently, electron microscopy and digestion experiments on nucleosome arrays gave new strong evidence for the zig-zag arrangement of the nucleosomes [31]. The x-ray structure of a compact tetranucleosome at 9 Å resolution has recently been reported by Schalch et al. [32], showing a zig-zag path of the DNA between the nucleosomes, which is not compatible with the one-start solenoidal helix. Geometric models build in this group on the base of the solved tetranucleosme resemble the crossed-linker model most closely.

Nevertheless the internal structure of the dense 30 nm chromatin fiber is still unclear. But for understanding gene regulation and epigenetics, the structure of chromatin and its changes as it folds into interphase and metaphase chromosomes are of fundamental importance. The essential mechanisms of condensation, decondensation and remodeling of chromatin involve stretching, bending and twisting of the 30 nm chromatin fiber and therefore depend on its flexibility.

The flexibility of a polymer chain can be expressed by its bending persistence length $L_p$, which is defined as the correlation length of the tangent vector to the chain axis. For the chromatin fiber, estimates of $L_p$ are controversial. Experimental, theoretical and simulation data support values over a wide range from 30 nm to 260 nm: Small values of $L_p$ = 30-50 nm are reported from scanning force microscopy (SFM) analysis of end-to-end distances of chromatin fibers on mica surface (Castro et al. [33] as cited by Houchmanzadeh et al. [34]). However, persistence lengths measured by SFM strongly depend on the binding conditions of the fiber to the mica [35]. Using optical tweezers to stretch chromatin fibers at low salt concentrations suggest $L_p$ = 30 nm [6], but no data for physiological salt was given there. Small persistence lengths of 30 – 50 nm were also postulated from recombination frequencies in human cells [36] and formaldehyde cross-linking probabilities in yeast [37].

Other groups report stiffer fibers with a persistence length in the range of 100-200 nm based on distance distributions for genetic markers pairs in human fibroblast nuclei [38-40] or recent experiments in budding yeast using *in situ* hybridization and live imaging techniques [41]. Stiffer fibers in the range of 200-250 nm are also support by computer simulations by Mergell and Schiessel [42].

Recent experiments by Cui and Bustamate [6], Bennink *et al.* [7], Brower-Toland *et al.* [8], and Leuba and Zlatanova *et al.* [43] investigated the mechanical properties of the chromatin fiber by single molecule stretching techniques. For forces below 10-20 pN, the extension of the chromatin chain is defined by its elasticity and no structural transition occurs, whereas forces above 10-20 pN lead to the disintegration of nucleosomes. Nevertheless quantities like the stretching modulus of a chromatin fiber are still unclear. Stretching a nucleosome-assembled lambda-phage DNA extract with an optical tweezers, Bennink *et al.* [7] derived a stretching modulus of 150pN for a salt concentration of 150 mM NaCl. Computer simulations suggested elastic moduli over a wide range from 60 to 240 pN depending on the salt concentration and the fiber geometry [42].

Here we used our earlier Monte Carlo (MC) model of the 30 nm chromatin fiber to model the elastic properties of compacted and stretched oligonucleosome chains. Before presenting our results, we shall first give a summary of the model and the basic methods for the computation of the persistence length and stretching modulus from our simulations.

# Methods

The model for simulating chromatin fiber stretching is based on a Metropolis Monte Carlo algorithm [44] and has been described earlier by Wedemann and Langowski [45]. It is used to generate a statistical relevant set of configurations representing the chromatin fiber at thermodynamic equilibrium at a defined temperature T. From the simulated conformational ensembles, we then extract the bending persistence length $L_p$ and the stretching modulus ε as fundamental mechanical properties of the 30 nm chromatin fiber.

### The chromatin fiber model

The chromatin fiber is approximated as a flexible polymer chain consisting of rigid ellipsoidal disks, 11 nm in diameter and 5.5 nm in height, which represent the nucleosome shape according to the crystal structure[11] (Fig. 1). In correspondence to the crystal structure the disks are connected by linker DNA, represented by two cylindrical segments. Incoming and outgoing linker DNA are set 3.1 nm apart of each other. The length of the linker DNA depends on the presence of linker histones and on the repeat length, which varies from organism to organism [12]. To explore the influence of the linker histone [26], simulations are performed with and without a stem motif added to each nucleosome. This is done by enhancing the distance between the nucleosome center and the attaching point of the linker DNA from 5.5 nm to 8 nm [26]. Knowing from X-ray crystallography [11] that 146 bp of DNA are wrapped around the histone octamer 1.65 times, our model of the nucleosome without a stem contains 2 full turns of DNA corresponding to 177 bp. Assuming a straight line of 22 bp from the exit point of the core particle to the end of the stem, leads to 146 bp + 2 x 22 bp = 190 bp of DNA associated to the nucleosome with stem. For instance, rat liver chromatin with a repeat of 200 bp [12] corresponds in our model to a free linker DNA of 10 bp if the linker histone is present, and 23 if absent.

The geometry used is essentially the "two angle" model developed by Woodcock et al. [22] and van Holde and Zlatanova [20], where the overall shape of the fiber is determined by the DNA-linker length, the linker DNA opening angle and the twisting angle between consecutive nucleosomes (Fig. 1). The stretching and torsion potentials of the joints connecting the linker DNA segments with each other and with the nucleosomes are assumed to be harmonic. The electrostatic DNA-DNA interaction is described by a Debye-Hückel approximation, while the nucleosome-nucleosome interaction is based on a weakly attractive anisotropic Lennard-Jones type (Gay-Berne-) potential. For an detailed description of our model see Wedemann et al [45].

**The stretching potential**

To simulate the stretching of the fiber, we added a constant force in x-direction. This is done by adding a pulling energy term $E_{pull}$ to the total energy of the conformations during the MC steps, which is proportional to the x-component of the mutual distance of the first and the last nucleosome of the fiber, $E_{pull} = -M \cdot \left| \vec{r}_{1,x} - \vec{r}_{N,x} \right|$, where M is the force modulus and $\vec{r}_{i,x}$ is the x-component of the position vector of nucleosome i.

**The simulation protocol**

The Monte Carlo scheme we use in our program consists of a pivot and a rotation move [46, 47]. In the pivot move, the shorter part of the chain is rotated by a random angle around a random axis passing through a randomly selected segment point. For the rotation move, the end point of a randomly chosen segment is rotated by a random angle around the axis determined by the start point of the chosen segment and the end point of the next segment.

The simulated chains consist of 100 nucleosomes. As in our previous simulations [45], the linker DNA entry-exit angle $\alpha$ is taken as 26° for the initial conformation. This value converges to an effective angle $\alpha_{sim}$ in the range of experimental values between 35° and 45° [26] due to the electrostatic forces and thermal fluctuations (Fig. 2).

In order to ensure the independence from the starting conformation we performed simulations for two energetically different starting conformations. In the first simulation run the starting conformation was a condensed fiber, where the sum of elastic energies is zero (Fig. 3.1). In the second simulation run we used an initial conformation where all segments are ordered in a straight line (Fig. 3.2)

The systems were then equilibrated. To check the statistics of the simulations, we calculated the autocorrelation function of the energy, end-to-end distance and mass density of the fibers for both simulation runs. The autocorrelation function of a quantity X calculated from the trajectory with respect to the number of simulation steps N is defined as $G(\Delta N) = \left\langle X(N) X(N+\Delta N) \right\rangle / \left\langle X(N)^2 \right\rangle$. During the MC procedure, G(ΔN) decreases exponentially with a typical 'correlation length' $N_{corr}$. We consider two conformations statistically independent if they are separated by at least $N_{corr}$ steps on the trajectory. For the relaxation of the total energy of both systems, we found a maximum of $N_{corr} \approx 3600$ MC steps, for the end to end distance $N_{corr} \approx 3200$ and for the mass density $N_{corr} \approx 2600$. Thus, we performed $5 \cdot 10^5$ MC steps for the initial relaxation of the chain corresponding to more than 100 statistically independent conformations.

Persistence length, diameter, mass density and total energy after equilibration agreed for both starting conformations (data not shown). After the equilibration the stretching potential was switched on and at least $3 \cdot 10^6$ MC steps were performed. For the final analysis, every 1000th conformation was used.

**The contour length**

In order to calculate the contour length $L_0$ of the 30 nm fiber, we first determined its axis. Our algorithm calculates subsequently the centers of mass $\vec{c}_i = \frac{1}{N_c} \cdot \sum_{j=i}^{i+N_c-1} \vec{R}_j$ , i = 1 .. $N$-$N_c$+1 of $N_c$ nucleosomes, where $\vec{R}_j$ is the position of nucleosome j. $N_c$ is called the nucleosome window length.

Connecting the centers of mass $\vec{c}_k$ results in a segmented chain describing the contour of the fiber. Its length $L_0 = \sum_{i=1}^{N-Nc} |\vec{c}_{i+1} - \vec{c}_i|$ is taken as the contour length and is sensitive to the window length. A large value of $N_c$ tends to average out the bending fluctuations within i and i+$N_c$, such that the contour length $L_0$ is underestimated. For a too small value of $N_c$ , the fiber axis follows the helical path of the linker DNA resulting in an overestimation of the contour length $L_0$.

To minimize this systematic error we performed simulations for different window sizes. We found that the mass density, elastic modulus and the persistence length plotted over the window length showed a plateau between a window of 6 and 22 nucleosomes (Fig. 4). Therefore we used the mean value of $N_c = 14$ for the data analysis.

**The persistence length $L_p$**

The persistence length $L_p$ is defined as the distance over which the direction of a polymer chain persists: the correlation between the orientations of two polymer segments decreases exponentially (with decay length $L_p$) with the contour length s separating them [48]. Since the thermal motions of a system are directly related to its stiffness, the persistence length $L_p$ of a chromatin fiber can be estimated directly from its fluctuations. Here, we calculated $L_p$ via two different routes, either from the local direction of the fiber axis given by the tangent to the contour or from the end-to-end distance of the fiber.

*Method A1: Fluctuations of tangent angles*

The decay of the correlation between the orientation of two fiber segments is given by the equation [49]:

$$\langle \cos(\Theta(s)) \rangle = \langle \vec{t}_i \cdot \vec{t}_j \rangle = \exp\left(-\frac{s}{L_p}\right) \qquad (1),$$

where s is the contour length separating the segments. Since the contour of the fiber is a segmented chain constructed from the centers of mass $c_i$, approximations to the tangent vectors $\vec{t}$ of the fiber axis can be calculated as $\vec{t}_i = \frac{\vec{c}_i - \vec{c}_{i+1}}{|\vec{c}_i - \vec{c}_{i+1}|}$, i = 1 .. N-$N_c$. Fitting an exponential decay to the autocorrelation function $\langle \vec{t}_i \cdot \vec{t}_j \rangle$ yields its correlation length which by definition equals $L_p$ .

*Method A2: Mean squared end-to-end distance*

The mean squared end-to-end distance for a flexible, but inextensible, worm-like chain depends on its contour length $L_0$ and its persistence length $L_p$ through the Kratky-Porod equation (2a) [50]:

$$\langle R^2 \rangle = 2 L_p^2 \left( \frac{L_0}{L_p} - 1 + \exp\left(-\frac{L_0}{L_P}\right)\right) \qquad (2a)$$

Equation (2a) describes the entropic elasticity of the worm-like chain arising from the reduced entropy of the stretched chain. Considering the entropic and enthalpic component as independent, the addition of the term $L_0 \cdot \frac{k_B T}{\varepsilon}$, with ε the stretching modulus of the fiber, yields to a similar equation for an extensible worm-like chain:

$$\langle R^2 \rangle = 2 L_p^2 \left( \frac{L_0}{L_p} - 1 + \exp\left(-\frac{L_0}{L_P}\right)\right) + L_0 \cdot \frac{k_B T}{\varepsilon} \qquad (2b)$$

From these relationships, the persistence length $L_p$ can be obtained for a given contour length $L_0$ from the mean squared end-to-end distance $\langle R^2 \rangle$.

**The stretching modulus ε**

The stretching stiffness of a fiber is given by the stretching modulus ε:

$\varepsilon := Y \cdot A$ , where *Y* is the Young's modulus of the material and *A* is the cross-sectional area.

We use two different methods for the calculation of the stretching modulus ε.

*Method B1: Fitting Hooke's Law*

The length of a polymer chain like the chromatin fiber depends on the bending angles of succeeding segments. Considering each angle as a random variable the central limit theorem of large numbers implies that for a sufficient large number experiments the distribution of the sum of these random variables converges against a normal distribution, resulting in low orders in a harmonic potential.

For a polymer chain the energetic form of Hooke's Law can be written as:

$$E_{str} = \frac{1}{2}\frac{Y \cdot A}{L_0}(L - L_0)^2 = \frac{1}{2}\frac{\varepsilon}{L_0}(L - L_0)^2 = \frac{1}{2}D(L - L_0)^2 \qquad (3),$$

where Y is the Young's modulus, A is the cross-sectional area, ε is the stretching modulus and $D := \frac{\varepsilon}{L_0}$ the force constant.

The contour length $L_0$ at no applied stretching force can be calculated directly from the simulations. The force constant D is obtained by fitting a parabola to the energy-extension curves in order to calculate the stretching modulus $\varepsilon = D \cdot L_0$.

*Method B2: The theorem for the equipartition of energy per degree of freedom*

Another way of obtaining the mean stretching energy at thermal equilibrium is based on the equipartition theorem, which states that each degree of freedom contains 1/2 kT of thermal energy. Then, together with Hooke's Law (3) we have:

$$\langle E_{str} \rangle = \frac{1}{2}kT = \frac{1}{2}D\left\langle (L - L_0)^2 \right\rangle \qquad (4)$$

Since $L_0 := \langle L \rangle$

$$\left\langle (L - L_0)^2 \right\rangle = \left\langle (L - \langle L \rangle)^2 \right\rangle = \langle L^2 \rangle - \langle L \rangle^2 = \langle L^2 \rangle - L_0^2 \qquad (5)$$

the force constant D can be calculated from

$$D = \frac{kT}{\langle L^2 \rangle - L_0^2} \qquad (6).$$

As before we obtain the stretching modulus ε from equation (3).

A stretching modulus ε is also obtained form the fit of equation (2b) characterizing an extensible WLC, but the fitting turned out to be not sensitive enough to estimate the stretching modulus. Fits with a similar $\chi^2$ agreed within a few percent in their persistence length, although the stretching moduli varied over 3 decades.

## Results

To study the dependence of the global stiffness of the chromatin fiber on the underlying geometry of the zig-zag chain, we varied the linker length l, the nucleosome-nucleosome twisting angle β and the opening angle α. Former simulations in our group have already shown the impact of these parameters on important fiber quantities like the linear mass density [45]. The choice of the twisting angle in the range of 90° to 130° gave the best agreement with the experimental nucleosome mass density of about 6 nucleosomes per 11 nm [26, 30]. Motivated by naturally occurring nucleosome spacing as in yeast and Hela cells, we focused here on nucleosome repeats of 195 – 205 bp [12], corresponding to a linker length of 5 – 15 bp. Since the attachment of the DNA to the nucleosome starts with the minor groove turned towards the first nucleosome binding site, the twisting angle β is a periodic function of the linker length [11, 51]: 1 helical turn of dsDNA corresponds to a length of 10.5 bp . However, for technical reasons and in order to study the effects more directly, we first varied the twisting angle β and linker length l independently.

For a detailed motivation of the elastic and interaction potentials see Wedemann et al [45]. A summary of initial parameters is presented in Table 1. Fig. 3 demonstrates a typical simulation run: Starting with either initial conformation (Fig. 3.1, Fig 3.2) the system is first equilibrated. In a second step the stretching force is applied to the relaxed conformation (Fig. 3.3) and the fiber is equilibrated again (Fig. 3.4 and Fig. 3.5).

### The bending stiffness of the fiber: the persistence length

We determined the persistence length for different linker lengths l and twisting angles β at an opening angle $\alpha_{init}$ = 26°, through the fluctuation of the tangent angles (A1) and the mean squared end-to-end distance (A2). Fig.5.1 and Fig 5.2 compare the results of the different analysis methods for simulations with and without a stem. The method based on the fluctuations of the tangent angles (A1) yields the lowest values for the persistence length. As shown in a typical fit to the data (Fig. 6) the fitted curve has a systematically lower decay length compared to the data. For short contour lengths, the segments are too long to reflect the exponential decay of a WLC (plateau in Fig. 6). Exclusion of first 20 data points from the fit of formula (1) yields a persistence length systematically lower by about 6 % (3%) for simulations with (without) stem. The values for the persistence length obtained from the fit of an extensible WLC to the mean squared end-to-end distance (A2) over contour length data are systematically 10 % (8 %) smaller than the values form the fit of an inextensible WLC with stem (no stem) and 5% (3 %) smaller than values from the numerical solution of (2a). The extensible WLC describes the flexibility of a fiber more realistically than the inextensible WLC, and since the additional term improves the fit by shifting the fitting graph along the y axis, a slightly softer fiber corresponding to a lower persistence length is obtained (Fig. 7). This persistence length obtained from the extensible WLC is in excellent agreement with the values from the tangent angles (A1) with a relative error smaller than 5 % (3 %) for simulations with (without) stem.

All applied analysis methods show the same trends for simulations with and without stem. An increasing linker length l reduces the persistence length and causes the fibers to behave softer, whereas fibers with higher twisting angles ß become stiffer and harder to bend as the increase in their persistence length suggests. For the further analysis of the persistence length we applied the method of fitting the extensible WLC (A2, 2b), which agrees with all other methods within less than 10 %.

Figure 8 shows the dependence of the persistence length on the initial opening angle $\alpha_{init}$ in the range form 22° to 34° for different sets of linker lengths and twisting angles. The equilibrium opening angle $\alpha_{eff}$ has a monotonic dependence on $\alpha_{init}$ as can be seen in Fig. 2. The results show that increasing the opening angle leads to a decrease in the persistence length, and as a result to a softer fiber. This trend is more pronounced for short linker lengths where as for longer linker lengths the bending stiffness is nearly constant. Since the electrostatic and nucleosomal interactions play a more important role for fibers with short linkers where nucleosomes approach each other more closely, such a behavior could be expected.

**The stretching stiffness of the fiber: the stretching modulus ε**

By fitting the energetic form of Hooke's Law to the data of the stretching simulations, we obtain the force constant D, resp. the stretching modulus ε for a given chain as presented in Fig 9.1. For this fit we have to ensure that the total or free energy, which includes both energetic and entropic contribution, can be set equal to the potential energy.

Therefore we first plotted the applied stretching force versus the end-to-end-distance of the stretched chromatin fiber. The values show an excellent linear behavior in the applied force region (Fig 9.2) which only holds if the entropic part can be neglected. The plot also suggests that for a force region smaller than 5pN the behavior of the chromatin fiber is non-linear which is in agreement with recent stretching simulation data [7].

Furthermore we calculated the stretching modulus from the force-contour length curve by fitting the linear function $F = \frac{\varepsilon}{L_0} \cdot L - \varepsilon$ to the data (Fig 9.3). Since this function is connected via an integral to method B1, the entropic part is also neglected. Fig 10.1 and 10.2 compare these results to the stretching moduli calculated form the theorem of equipartition (B2), which includes both energetic and entropic part of the energy. The mean difference of the values is smaller than 10 % (13 %) for simulations without (with) stem and supports the negligence of the entropic part. Fig. 11.1 and 11.2 present the results of the method using the Hooke's Law (B1), which show a good agreement with the elastic moduli estimated by method applying the theorem of equipartition (B2), except for high twisting angles and low linker lengths. Calculating the stretching moduli ε by method B2 for all stretching simulations (0, 5, 10, 15, 20 pN) and taking the mean value as stretching modulus, we could improved the accuracy and reduce the relative difference between both methods by 18 % (8%) for simulations with stem (without stem) as shown in Fig. 11. In spite of the differences in the absolute values for the stretching modulus for high twisting angles and low linker lengths all methods show the same trends as already described for the persistence length. With increasing linker lengths l and opening angle α between the nucleosomes, the elastic moduli decrease and the fiber becomes softer. Increasing the twisting angle β leads to a rise of the elastic moduli and the fiber becomes harder to stretch.

**How stiff is the 30 nm chromatin fiber ?**

Both stretching and bending stiffness of the simulated chromatin fiber show the same qualitative behavior: the stiffness increases with nucleosome twisting angle and decreases with linker length.

In order to compare directly bending and stretching stiffness, we first assume that the 30 nm chromatin fiber behaves as a homogenous isotropic elastic rod with circular cross-section. Under this condition the persistence length $L_p$ can be derived from the force constant D as follows:

Bending two segments of a polymer chain located a distance L apart by an angle θ requires the energy [52]

$$E_{bend} = Y \cdot I \cdot \Theta^2 / 2L \qquad (7),$$

where Y is the modulus of extension or Young's modulus and I is the cross section moment of inertia of the molecule.

These macroscopic quantities are related to the persistence length of the polymer $L_p$ by

$$L_p = \frac{Y \cdot I}{kT} \qquad (8)$$

where k is the Boltzmann constant and T the absolute temperature.

Solving Hooke's Law (3) for Y and substituting in (8) results in:

$$L_p = \frac{1}{kT} \frac{D \cdot L_0}{A} \cdot I \qquad (9)$$

With the cross section moment of inertia I of a homogeneous elastic rod

$$I = \frac{1}{4}\pi R^2 \qquad (10)$$

we obtain the final expression for the persistence length $L_p$:

$$L_p = \frac{1}{4}\frac{DL_0}{kT}R^2 \; . \qquad (11)$$

We now compare the persistence lengths $L_{p,\mathrm{B}}$ derived from the force constant D under the assumption of a homogeneous isotropic elastic rod with circular cross-section to the persistence lengths extracted directly from the bending fluctuations. The results are shown in Fig. 12.1-12.4 and demonstrate clearly that the persistence length $L_{p,B1}$ expected from a homogeneous elastic rod is 6-10 times higher than the directly extracted persistence length $L_{p,A1}$ under all conditions for both simulations with and without stem. Thus the chromatin fiber is much easier to bend than expected for a homogenous elastic rod; its stretching rigidity is higher than its bending rigidity. Single molecule experiments show a similar behavior for DNA [53].

**Effect of the nucleosome stem**

In order to explore the influence of the linker histone-induced stem motif on the persistence length of the chromatin fiber, simulations with the same repeat length have been compared with and without stem. For these simulations the twisting angle β is adjusted to its corresponding linker length according to experimental data [11, 51]: 1 helical turn of dsDNA corresponds to a length of 10.5 bp. Figure 13.1 and 13.2 present the results of these simulations for a range of repeat lengths form 191 bp to 220 bp. For all repeat lengths in the simulations with and without a stem the persistence length shows a general trend: with increasing repeat length the persistence length decreases and the fiber becomes softer. This agrees with the data shown in Fig. 12.1 and 12.2 in which the twisting angle is set constant. The trend is more pronounced for short repeat lengths since the electrostatic and internucleosomal forces are strong at short distances. But peaks in figures 13.1 and 13.2 also show that softening of the fibers due to a higher linker length can be overcome by a high torsion angle, which is known to stiffen the fiber as we showed in Figure 12.1 and 12.2 before. This effect is only dominant at short repeat lengths and loses its influence as the interactions become weaker for longer repeat lengths. The occurrence of such peaks could be one reason for the preference for certain repeat lengths of the chromatin fiber in organisms. Geometric models based on the crystal structure of the nucleosome and ideal B-DNA, which explicitly account for sterical hindrance, also suggest such preferred conformations (S. Diekmann, personal communication). Fig 13.1 and 13.2 clearly show that the stiffness of the fibers with stem is higher than for fibers lacking the stem, although this difference becomes smaller for longer repeat lengths. This might by a hint for the stabilizing role of a stem as already suggested from experimental data [54].

**Discussion**

Our results show that the bending and the stretching stiffness of the chromatin fiber strongly depend on the local geometry of the nucleosome. Both the persistence length $L_p$, characterizing the bending stiffness of the fiber, and the stretching modulus $\varepsilon$, which describes the stretching stiffness of the fiber, decrease if either the linker lengths are increased from 5 to 15 bp, the opening angle is increased from 22° to 24° or the twisting angle is reduced form 130° to 90°. This behavior is independent of the presence of a stem motif, which models the linker histone. A linker histone is known to decrease the opening angle $\alpha$ between the entry and exit of the linker DNA and as a result leads for high salt concentrations to a more condensed fiber structure [54]. This is supported by our simulations since the presence of a stem motif provides higher persistence lengths thus softer fibers (Fig. 13.1, 13.2).

Our major result is based on the comparison of the stretching and bending stiffness assuming chromatin to behave as a homogenous isotropic elastic rod with circular cross-section, which showed that the chromatin fiber is more resistant to stretching than to bending. This property of the chromatin fiber is important for its ability to condense and decondense, for example to prevent or allow transcriptional access. Interpreting our result from a higher level of compaction, for chromatin fibers it seems more favorable to be packed via dense loops than by a linear compression. The formation of hairpin structures has been observed in cryo-EM pictures under the presence of MENT, a heterochromatin protein that mediates higher order chromatin folding [55]. Some hairpin conformations could also be seen in our simulations.

As far as the ratio of stretching and bending elasticity is concerned, single molecule stretching experiments [53] suggest that dsDNA behaves different from chromatin: The stretching modulus $\varepsilon$ of dsDNA for physiological salt conditions is estimated to ~1100 pN [56, 57]. Applying (11) under the assumption of a homogenous elastic rod with a radius of 1 nm to dsDNA yields a bending persistence length of $L_p$ = 70 nm, which is about a factor of 1.4 higher then the known persistence length of 53 nm for dsDNA [58]. Since dsDNA is not an isotropic cylinder and the effect of the grooves will decrease the cross-sectional moment of inertia somewhat, the effective radius may be taken as smaller than 1 nm. Since a reduction of the radius by 20 % leads to an agreement of the persistence length known from bending and of that estimated from the stretching modulus, we can state that dsDNA is almost equally resistant to stretching and to bending. For chromatin this difference is at least 4 times higher (Fig. 12). This behavior makes sense for its biological role, since the chromatin fiber could be easily opened to proteins by simple bending or torsion.

As mentioned in the introduction, the exact value for the persistence length of the chromatin fiber is still under discussion, with estimates ranging from 30 nm to 260 nm. Some of the small values in this range were obtained at low salt concentrations, where a smaller persistence length compared to our results for high salt can be expected, since low salt is known to open the fiber. Other experiments resulting in small persistence lengths were done in constrained volumes by cross-linking procedures[36, 37]. Under these circumstances, the condition of an unconstrained self-crossing walk is only fulfilled over short distances. Thus, for a given chain flexibility, the measured apparent persistence length will depend on the genomic separation and folding topology for which it is calculated [41]. Furthermore, a persistence length in the range of the fiber diameter of 30 nm would lead to extremely irregular structures, which are hard to be reconciled with the concept of a “fiber”.

Analysis of the distance distribution for genetic marker pairs in human fibroblast nuclei [38-40] provide higher values of $L_p$ = 100 – 140 nm based on a wormlike chain model. Recent experiments in budding yeast using optimized *in situ* hybridization and live imaging techniques [41] report stiff interphase chromatin fibers estimating a persistence length of 120-200 nm. These are supported by our simulations for short linker lengths of 5 bp and a stem which suggest $L_p$ = 140-220 nm (Table 2). The persistence lengths that we obtained for linker lengths of 10 and 15 bp are in the range 50 – 280 nm as shown in Table 2, decreasing with longer linkers. Similar persistence lengths between 200-250 nm are reported from computer simulations carried out recently by Mergell and Schiessel [42]. They followed our approach in modeling the nucleosome-nucleosome interaction as ellipsoids interacting via a Gay-Berne potential, but no displacement of the entering and exiting DNA at the nucleosome in the direction of the nucleosome axis was implemented. In addition our model has the advantage of explicit inclusion of the screened electrostatic interaction between the linkers via a Debye-Hückel potential that defines the opening angle $\alpha$ as a function of the salt concentration. Mergell and Schiessel [42] reported elastic moduli in the range of 60 to 240 pN, in agreement with our data (Table 3).

Recently Sun et al. using an irregular Discrete Surface Charge Optimization (DiSCO) model for their MC simulation reported data on nucleosomal arrays. They simulated fibers consisting of 12 nucleosomes, which is 12 % of the number of nucleosomes in our simulation, and focused on the salt-dependent condensation of the chromatin fiber. Unfortunately no data concerning persistence length and elastic modulus is given here.

Bennink et. al. derived 150 pN as stretching modulus for a salt concentration of 150 mM NaCl [7], using optical tweezers for the stretching of a nucleosome-assembled lambda-phage DNA extract of a *Xenopus laevis* egg with no linker histones attached and nucleosome repeat length of 200 bp. Our simulations yield a lower value of 40 pN already for a repeat length of 192 (no stem). One reason for this discrepancy may be the difference of 50 mM in the salt concentrations, since our simulation parameters have been calibrated for 100 mM NaCl. A lower salt concentration leads to a lower compaction thus to a lower stretching modulus. Furthermore the solution used in the tweezers experiment contains proteins known to act close to the entry-exit points similar to the linker histones. This is supported by our simulation, which for a repeat length of 200 bp (with stem) yields a stretching modulus in the range of 90-160 pN. Nevertheless the Gay-Berne potential used in our model is only an approximation of the nucleosome-nucleosome interaction, which plays a major role at physiological salt conditions. To improve the quantitative predictions of our model, more detailed interaction potentials are need, including for example the salt depended effects of the histone tails. In particular for the modeling of nucleosome unwrapping and interpreting corresponding experiments[7, 8], the DNA – nucleosome interaction is of great interest. Inclusion of such potentials into the chromatin fiber model will provide a deeper insight in the architecture and behavior of the chromatin fiber as it undergoes biologically important modifications, as well as into its role in transcription and gene regulation.

**Conclusion**

Based on the “two – angle” model of the 30 nm chromatin fiber, we performed computer simulations in order to estimate systematically two fundamental but still controversial physical parameters: the stretching modulus ε and the persistence length $L_p$. Our results show a strong dependence on the linker length, the torsion angle and the opening angle for both simulations with and without a stem. A rise of the linker DNA length from 5 to 15 bp leads to a softening of the fiber by at least a factor of 5. A similar but weaker effect has been observed for short linker lengths if the opening angle is increased from 22° to 34°. A significantly stiffer fiber was the result of increasing the twisting angle between nucleosomes form 90° to 130°. Furthermore a comparison of simulation data with stem to data without a stem for equal repeat lengths showed that fibers with stem are stiffer. Most importantly, we investigated the directional dependence of the chromatin fiber rigidity. We clearly showed that its rigidity depends on the direction of deformation. The chromatin fiber does not behave as an isotropic elastic rod, rather it is much more resistant to stretching than to bending. For a complete understanding of higher order chromatin folding, it will be important to include these nanomechanical parameters in models of chromosomes.

**Acknowledgements**

This work was supported by a grant of the VW foundation in the program ‘Physics, Chemistry and Biology with Single Molecules’.

**TABLE 1:** Elastic, interaction and geometric parameters used in the simulations.

| Parameter | Measure |
|---|---|
| Stretching modulus DNA | 1104 pN |
| Bending modulus DNA | $2.06 \cdot 10^{-19}$ J nm |
| Torsion modulus DNA | $2.67 \cdot 10^{-19}$ J nm |
| Electrostatic radius DNA | 1.2 nm |
| Stretching modulus nucleosome | 1104 pN |
| Torsion modulus nucleosome | $1.30 \cdot 10^{-19}$ J nm |
| Gay-Berne parameters for internucleosome Interaction | $\sigma_0 = 10.3$ nm<br>$\chi = -0.506$<br>$\chi' = -0.383$ |
| Temperature | 20 °C |
| Nucleosome diameter | 11 nm |
| Nucleosome height | 5.5 nm |
| Salt concentration | 0.1 M NaCl |

**TABLE 2:** Persistence length $L_{p,A1}$ calculated by the fluctuation of the tangent angles (A1) for different repeat lengths $l_{rep}$ and twisting angles β.

**Persistence length, no stem**

| $l_{rep}$ [bp] \ β [°] | **90** | **110** | **130** |
|---|---|---|---|
| **182** | 261.1 | 508.4 | 1090.0 |
| **185** | 177.1 | 309.3 | 546.7 |
| **187** | 125.8 | 183.2 | 280.0 |
| **190** | 76.2 | 95.9 | 146.0 |
| **192** | 53.9 | 56.0 | 69.8 |

**Persistence length, stem**

| $l_{rep}$ [bp] \ β [°] | **90** | **110** | **130** |
|---|---|---|---|
| **195** | 145.4 | 220.3 | 583.2 |
| **198** | 105.8 | 145.5 | 283.4 |
| **200** | 83.5 | 97.6 | 142.0 |
| **203** | 65.1 | 64.2 | 60.0 |
| **205** | 48.6 | 51.9 | 40.9 |

**TABLE 3:** Stretching modules ε calculated by fitting Hooke's Law s (B1) for different linker lengths l and twisting angles β.

| Stretching modulus<br>α = 26°, no stem<br>F= 0,5,10 pN | β = **90°**<br>ε [pN] | β = **110°**<br>ε [pN] | β = **130°**<br>ε [pN] |
|---|---|---|---|
| $l_{rep}$ = **182bp** | 321.07 | 656.10 | 1250.75 |
| $l_{rep}$ = **185 bp** | 181.32 | 263.16 | 590.47 |
| $l_{rep}$ = **187 bp** | 100.43 | 132.62 | 284.88 |
| $l_{rep}$ = **190 bp** | 59.64 | 78.65 | 116.33 |
| $l_{rep}$ = **192 bp** | 44.43 | 49.58 | 52.29 |

| Stretching modulus<br>α = 26° , stem<br>F= 0,5,10 pN | β = **90°**<br>ε [pN] | β = **110°**<br>ε [pN] | β = **130°**<br>ε [pN] |
|---|---|---|---|
| $l_{rep}$ = **195 bp** | 95.38 | 264.10 | 468.10 |
| $l_{rep}$ = **198 bp** | 73.97 | 151.69 | 276.20 |
| $l_{rep}$ = **200 bp** | 66.10 | 89.73 | 154.29 |
| $l_{rep}$ = **203 bp** | 58.45 | 68.18 | 73.23 |
| $l_{rep}$ = **205 bp** | 46.99 | 54.68 | 41.29 |

# Legends to Figures

**Fig. 1:** The geometry of the implemented fiber model. Incoming and outgoing linker DNA of length L build the opening angle $\alpha$ and are set a distance d off each other. Succeeding nucleosomes are twisted by the torsion angle $\beta$. Interactions between the linker segments are described by a Deby-Hückel-Potetial whereas nucleosomes interact via a Gay-Berne-Potential.

**Fig. 2:** Mean effective opening angle $\alpha_{eff}$ as a function of the initial opening angle $\alpha_{init}$ . During relaxation the initial opening angle $\alpha_{init}$ converges to the effective opening angle $\alpha_{eff}$ due to electrostatic repulsion and thermal fluctuations. $\alpha_{init}$ and $\alpha_{eff}$ show a monotonic and nearly a linear behavior.

**Fig. 3:** Stretching simulation of a fiber consisting of 100 nucleosomes (red), linker segments (blue) of repeat length l = 205 nm, a opening angle $\alpha_{init}$=26 ° and a twisting angle $\beta$=90°. Starting the relaxation of the chromatin fiber from a twisted (3.1) and linear (3.2) initial conformation the chromatin fiber is equilibrated. Fig. 3.3 is a typical example of an equilibrated structure. In a second step an external pulling force of $F_{pull}$ = 5 pN is applied and further equilibration initiated. The application of the stretching force leads to an extension of the fiber as shown in Fig 3.4 and Fig 3.5.

**Fig. 4:** Persistence length as function of the nucleosome window length for different linker lengths and torsion angles. All functions show a plateau between a window of 6 and 22 nucleosomes. dotted line : repeat length l = 190 bp, $\beta$=130°, no stem; fine line: repeat length l = 190 bp; $\beta$=110°, no stem; bold line: repeat length l = 190 bp; $\beta$= 90 normal line: repeat length l = 200 bp, $\beta$= 90°, stem °, no stem; dashed line: repeat length l = 200 bp, $\beta$=130°, stem; dash-dot line : repeat length l = 200 bp, $\beta$=110°, stem

**Fig. 5:** Persistence length $L_p$ calculated by the fluctuation of the tangent angles (A1, open symbols, solid lines) and the fluctuation of the squared end-to-end-distance (A2, closed symbols, dashed lines line) as a function of different linker lengths l and twisting angles $\beta$. All methods show the same general trend: with increasing linker length and decreasing torsion angles the persistence lengths of the fibers decrease and the fiber becomes easier to bend. The fit

of the extensible WLC (A2, 2b), which agrees with all other methods within less than 10 % and is used to calculate the persistence length $L_p$. ○: A2, inextensible WLC by numerical solution of (2a); ◇: A2, inextensible WLC by fitting (2a);
△: A2, extensible WLC by fitting (2b); ●: A1, fitting exponential decay (1)

**Fig. 6:** Fit of equation (1) describing the exponential decay of the correlation of the tangent angles to the data of a 100 nucleosome chromatin fiber with 187 bp, β = 110° and no stem. For the first data points the segments are too long compared to the contour length and as a result the correlation is unrealistically high. Thus these points are excluded from the fit and the peak cannot be reflected in the exponential fit. This fit also shows that the decay length of the fitted curve is slightly lower than can be expected from the data.
○: 187 bp, β = 110°, no stem; solid line: A1, fit of exponential decay (1)

**Fig. 7:** Fit of the Kratky-Porod equation (2a) describing an inextensible WLC and fit of equation (2b) describing an extensible WLC to the data of a 100 nucleosome chromatin chain with 187 bp repeat length, β = 110° and no stem. Both fits show an excellent agreement with the data. The $\chi^2$ of the extensible WLC is slightly lower, reflecting the more realistic description of the fiber.
○: 187 bp, β = 110°, no stem; ●: extensible WLC, eq. (2b); solid line: inextensible WLC, eq. (2a)

**Fig. 8:** The persistence length for different sets of repeat lengths and twisting angles as a function of the initial opening angle $\alpha_{init}$. An increase in the opening angle leads to a decrease in the persistence length corresponding to softer fibers. This tendency is stronger for short linker lengths, where electrostatic and nucleosomal interactions are more intense than for longer linker lengths. ○ : repeat length = 188, twisting angle β = 110°, no stem; ◇: repeat length = 192, twisting angle β = 120°, no stem; △: repeat length = 199, twisting angle β = 110°, no stem.

**Fig. 9.1:** Total energy of a 100 nucleosome chromatin chain for a repeat length of 190 bp, β = 110° and no stem as function of the chain contour length. The elastic force constant D is obtained by fitting Hooke's Law $E = \frac{1}{2} D (L - L_0)^2 + const$ to the data.

▬ : 0 pN ; ▬ : 5 pN ; ▬ : 10 pN ; ▬ : 15 pN ; ▬ : 20 pN

**Fig. 9.2:** Stretching force as a function of the end-to-end distance for a 100 nucleosome chromatin chain of 190 bp repeat length, β = 110° and no stem. The linear behavior shows that entropic contributions in the force range from 5 to 20 pN can be neglected. In the lower force range a non-linear behavior can be expected, which is in agreement with recent stretching experiments.

**Fig. 9.3:** Stretching force plotted versus the contour length of a 100 nucleosome chromatin fiber for a repeat of 190 bp, β = 110° and no stem. The stretching modulus ε is calculated from the fit of $F = \frac{\varepsilon}{L_0} \cdot L - \varepsilon$ to the data.

**Fig. 10:** Stretching modules ε calculated by a linear fit to the force-extension curve (open symbols, solid lines) compared to the method of using the theorem of equipartition (B2, closed symbols, dashed lines) for simulation runs without stem (Fig. 10.1) and with stem motif (Fig. 10.2) as a function of different linker lengths l and torsion angles β. Both methods show a good agreement, which supports the assumption that the entropic contribution to the total energy can be neglected in this force region. With increasing linker length and decreasing torsion angle β the stretching modules ε of the fibers decrease and the fiber becomes easier to stretch. This is true for both methods and independent of the presence of a stem. Circles: β = 130°; diamonds: β = 110°; triangles: β = 90°.

**Fig. 11:** Stretching modules ε calculated by fitting Hooke's Law (B1, open symbols, solid lines) and by using the theorem of equipartition (B2, filled symbols, dashed lines) for simulation runs without stem (Fig. 11.1) and with stem motif (Fig. 11.2) as a function of different linker lengths l and torsion angles β. Both methods show a good agreement except for small repeat lengths and high torsion angles. Independent of the presence of a stem the stretching modulus ε of the fibers decreases with increasing linker length l and decreasing torsion angle β in agreement with Fig 10. Symbols and lines as in Fig. 10.

**Fig. 12:** Examination of the directional dependence of the fiber stiffness by comparing the persistence lengths for different linker lengths l and twisting angles β. The solid lines represent the persistence lengths $L_{p,A1}$ calculated directly by fluctuation of the tangent angles (A1, open symbols, solid lines) whereas the dashed lines show the persistence lengths $L_{p,B1}$ derived under the assumption that the fiber behaves like a homogeneous isotropic elastic rod with circular cross-section using the fit of Hooke's Law to the stretching data (B1, closed symbols, dashed lines). $L_{p,B1}$ is 6-10 times higher compared to $L_{p,A1}$ for simulations without stem (Figs. 10.1, 10.3) and with stem (Figs. 10.2., 10.4)

Symbols in 12.1, 12.2: Circles: β = 130°, diamonds: β = 110°; triangles: β = 90°, in 12.3: circles: repeat length 188 bp, β = 110°, diamonds: repeat length 192 bp, β = 120°; triangles: repeat length 199 bp, β = 110°, in 12.4:

circles: repeat length 201 bp, β = 110°, diamonds: repeat length 205 bp, β = 120°; triangles: repeat length 212 bp, β = 110°.

**Fig. 13:** Examination of the stem influence for simulations with equal repeat length. For these simulations the twisting angle β is adjusted to its corresponding linker length: 1 helical turn of dsDNA corresponds to a length of 10.5 bp. Fig 13.1 and Fig.13.2 show that the persistence lengths of fibers with stem (closed circles, dashed lines) are higher than for fibers without stem (open squares, solid lines). This effect is stronger for short repeats and weakens with increasing repeat length. The peaks show that the twisting angle strongly influences the stiffness of the fiber and can lead to stiffer fibers although for longer linker lengths a softer fiber could be expected.

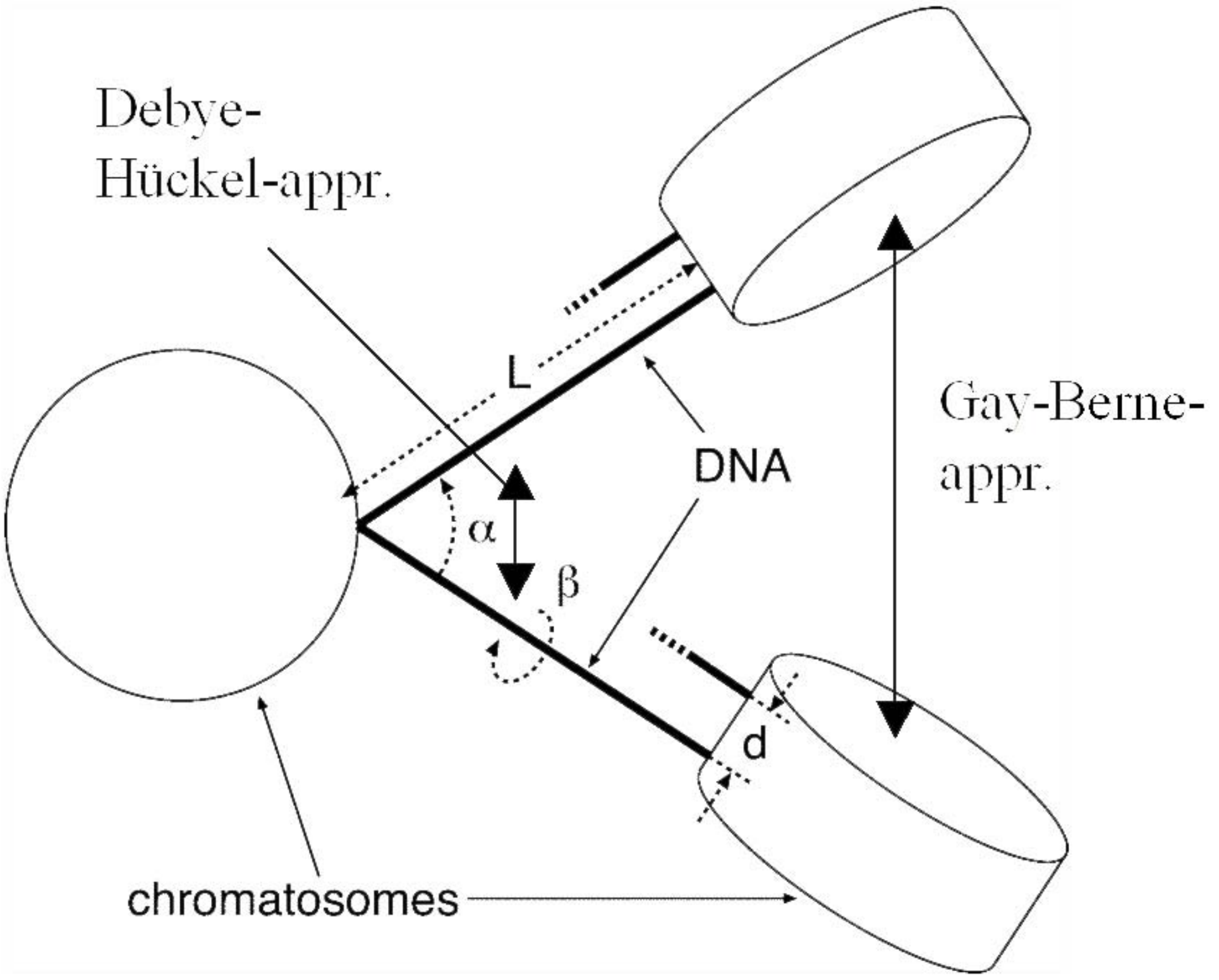

Fig. 1

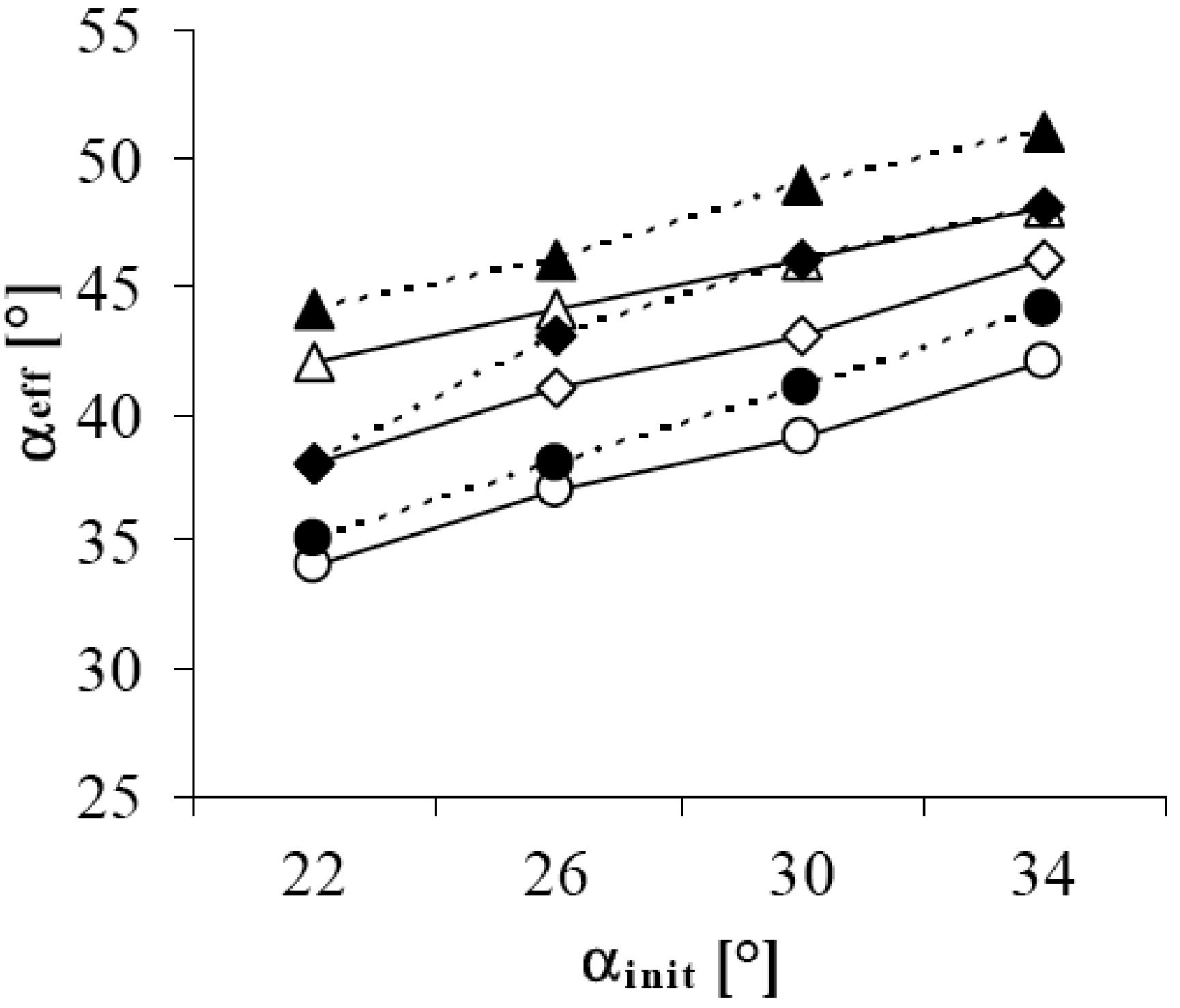

Fig. 2

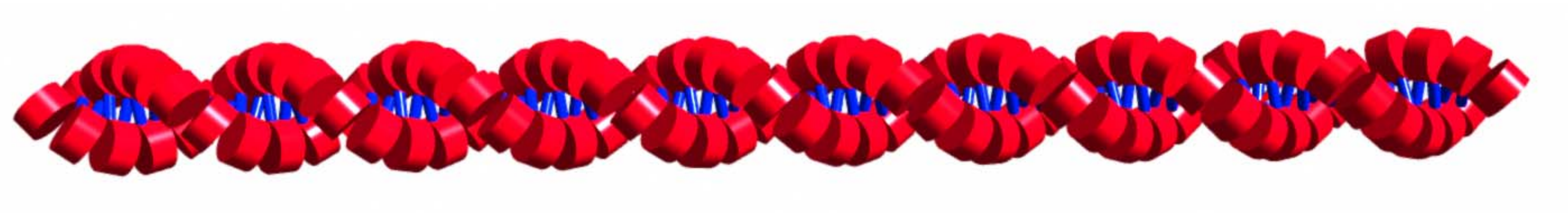

Fig. 3.1

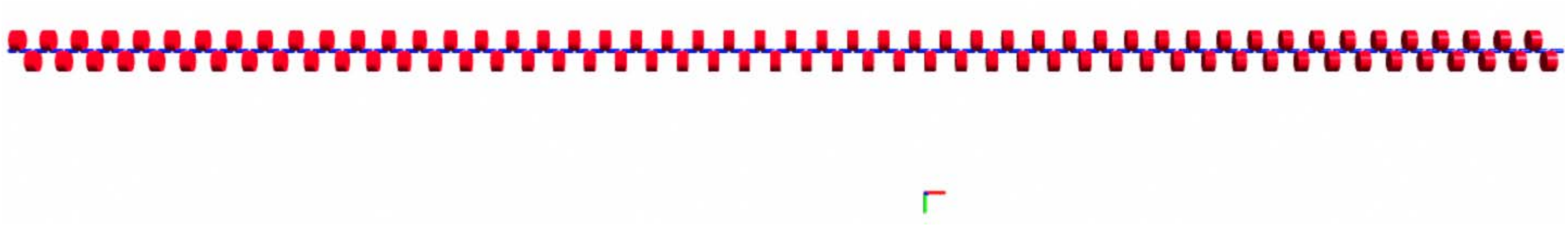

Fig. 3.2

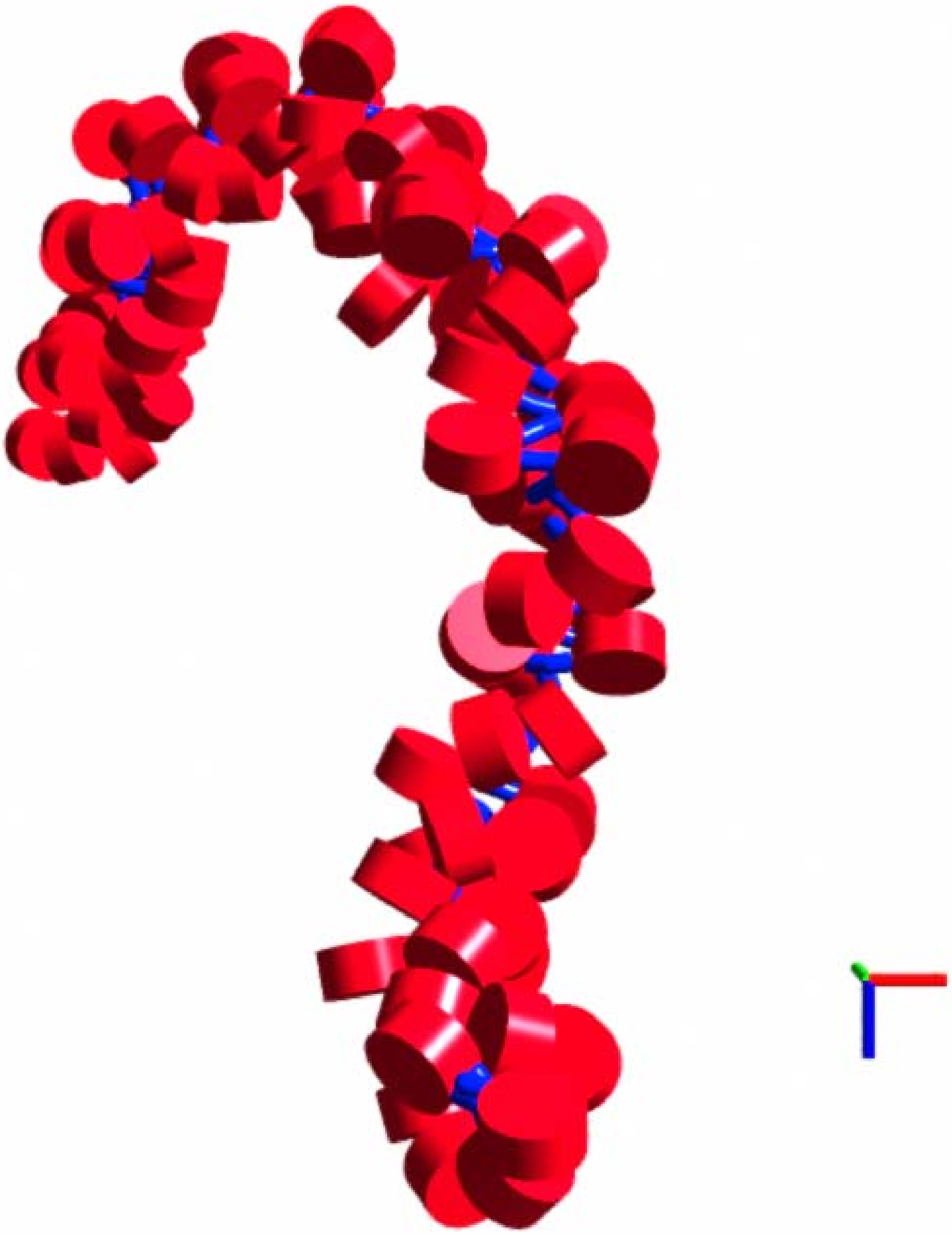

Fig. 3.3

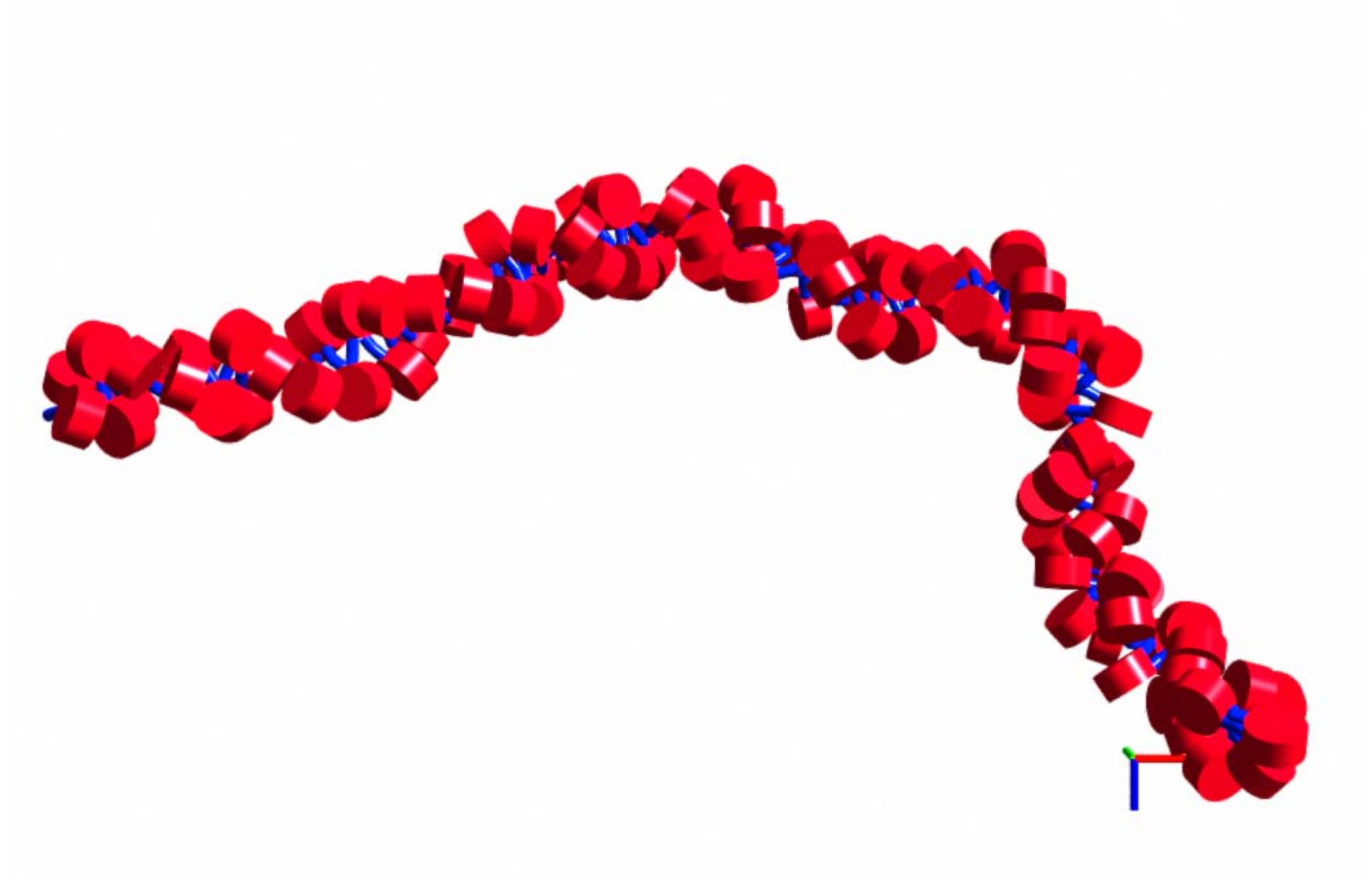

Fig. 3.4

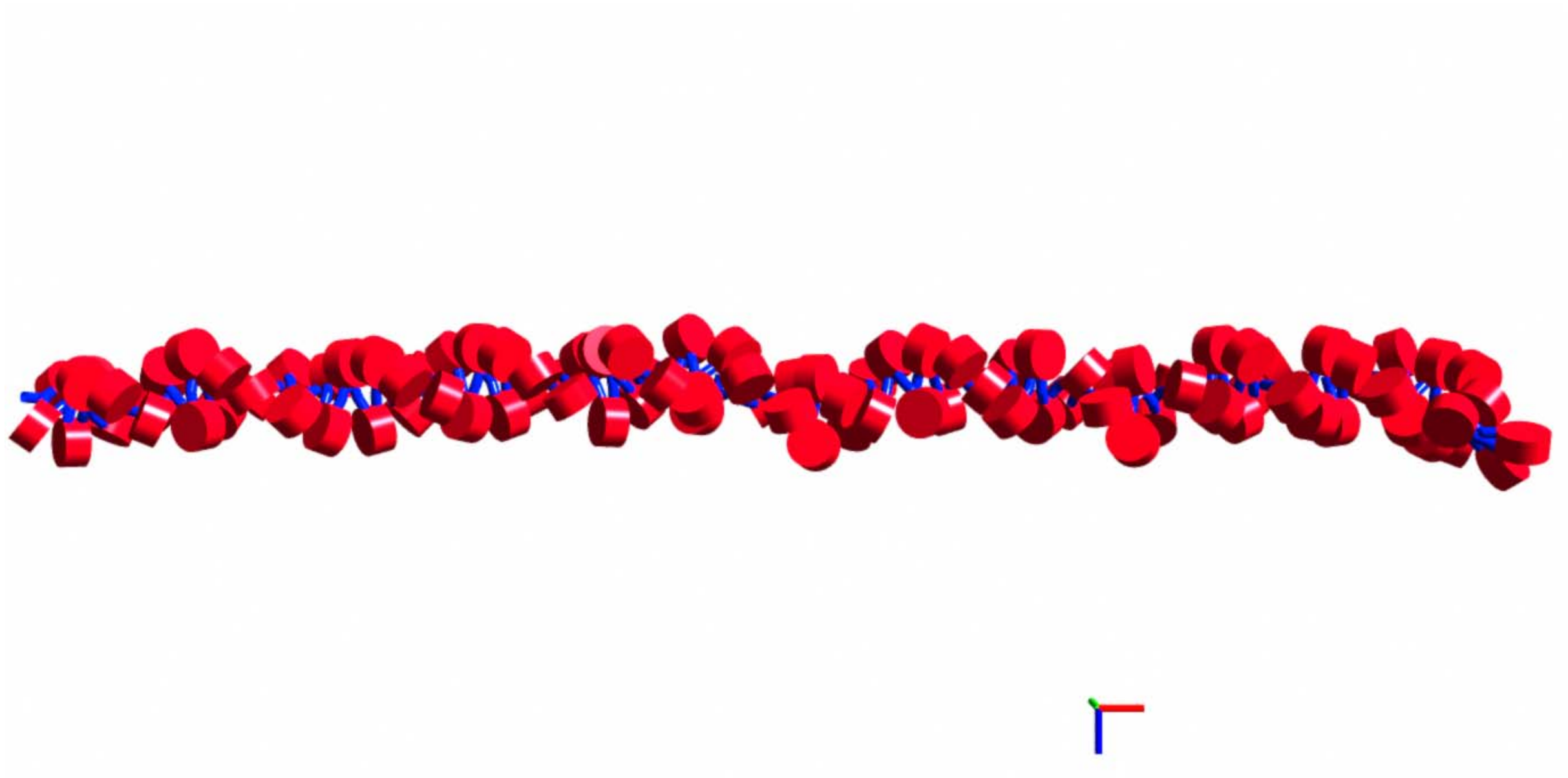

Fig. 3.5

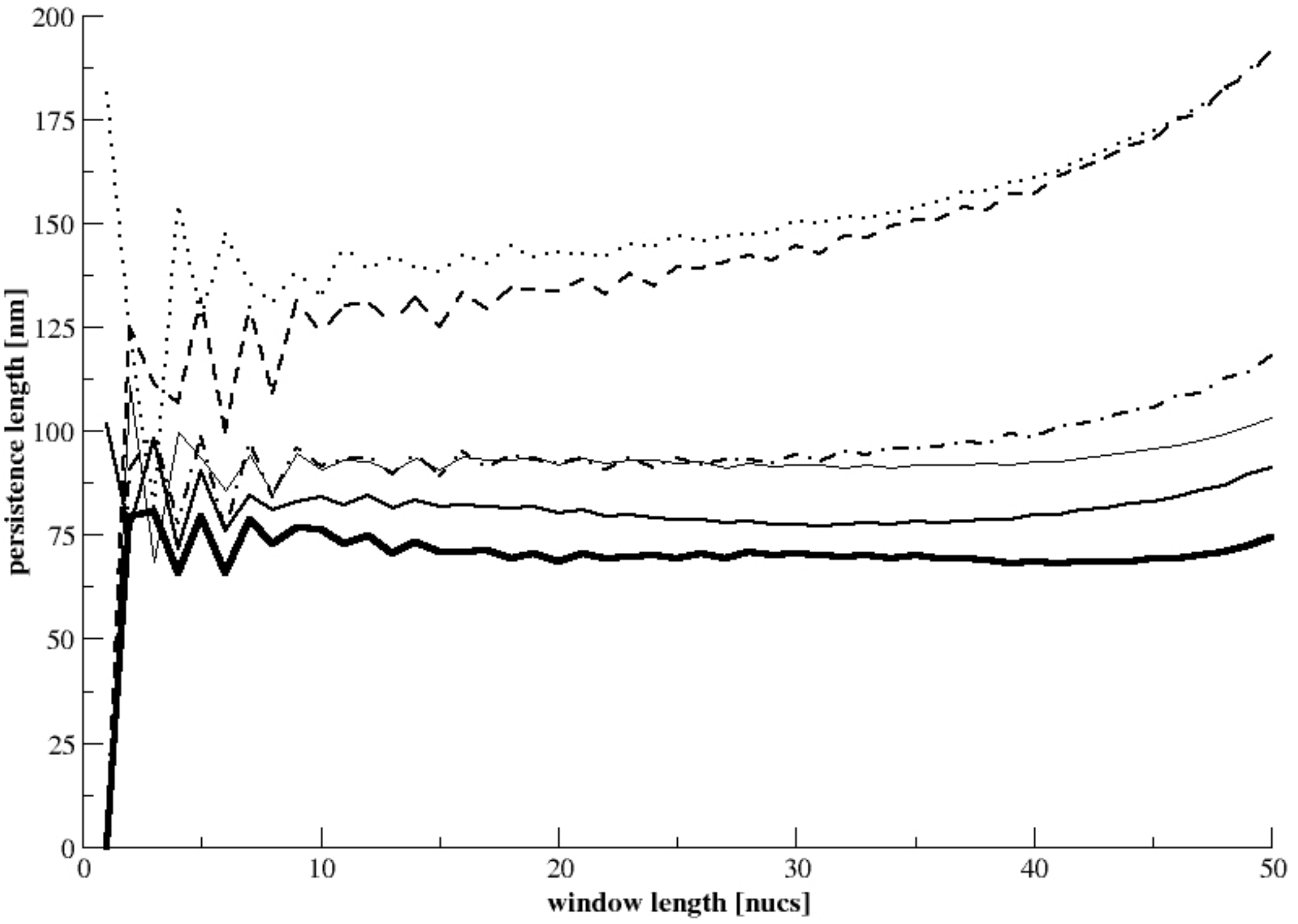

Fig. 4

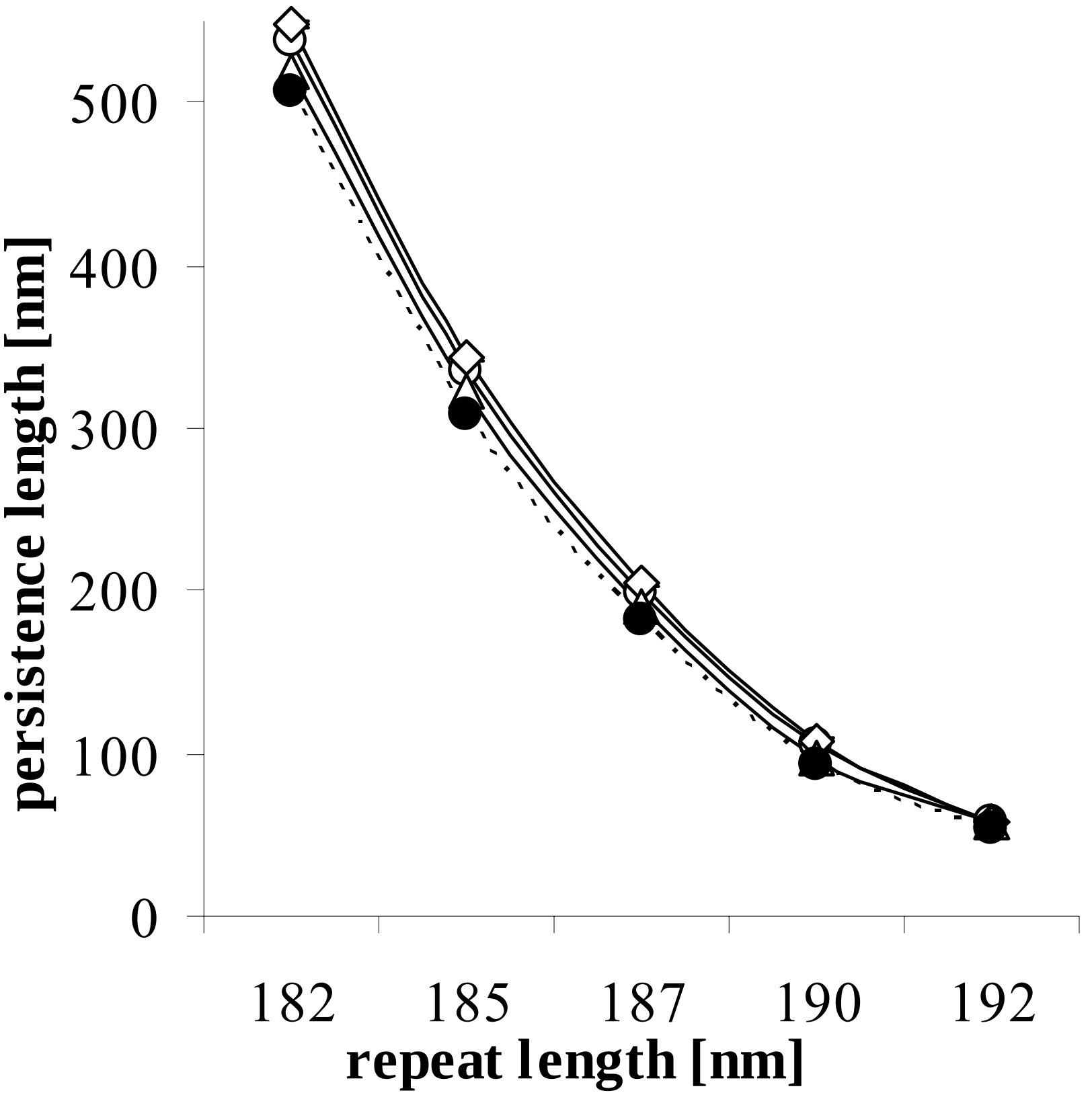

Fig. 5.1

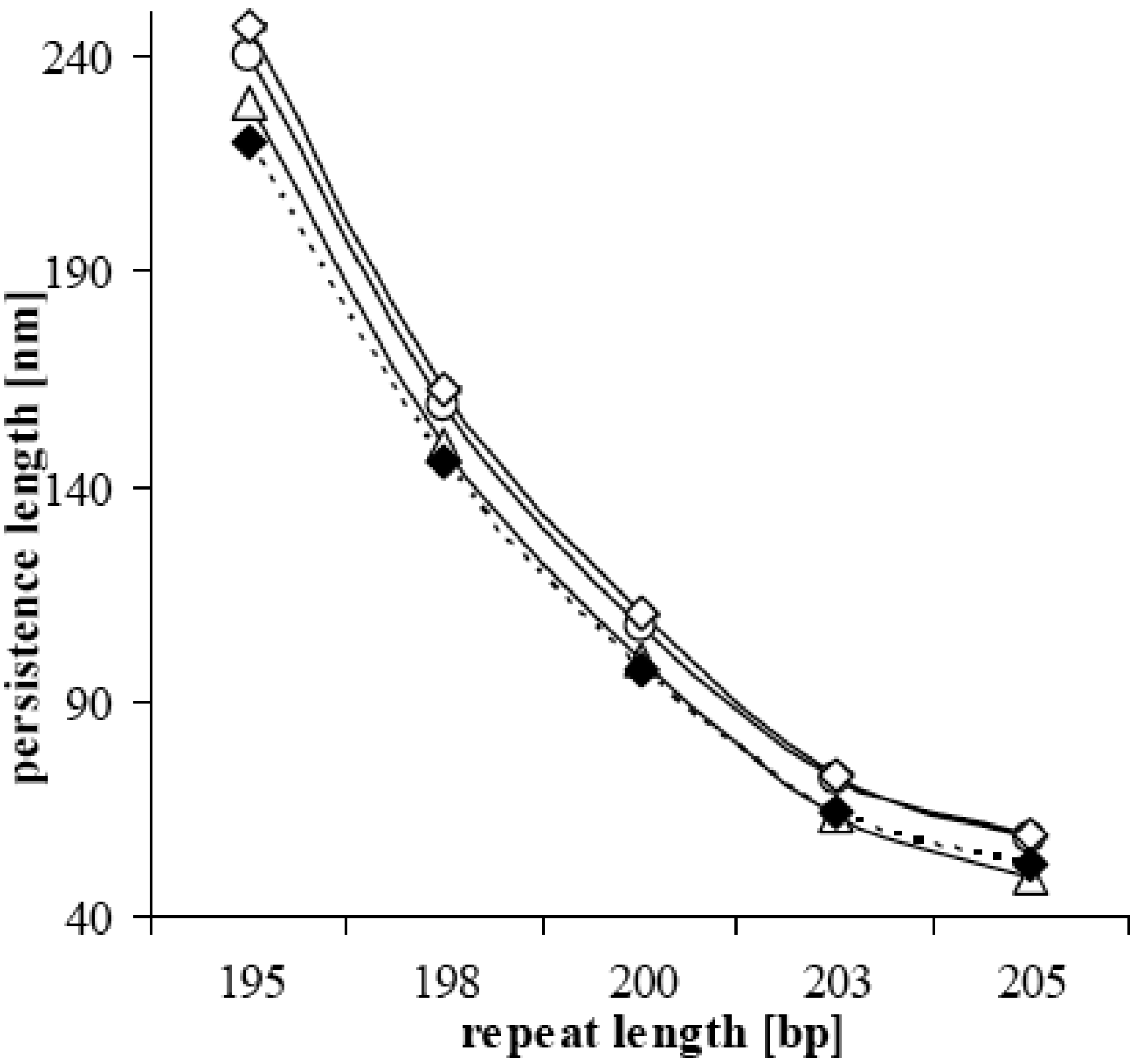

Fig. 5.2

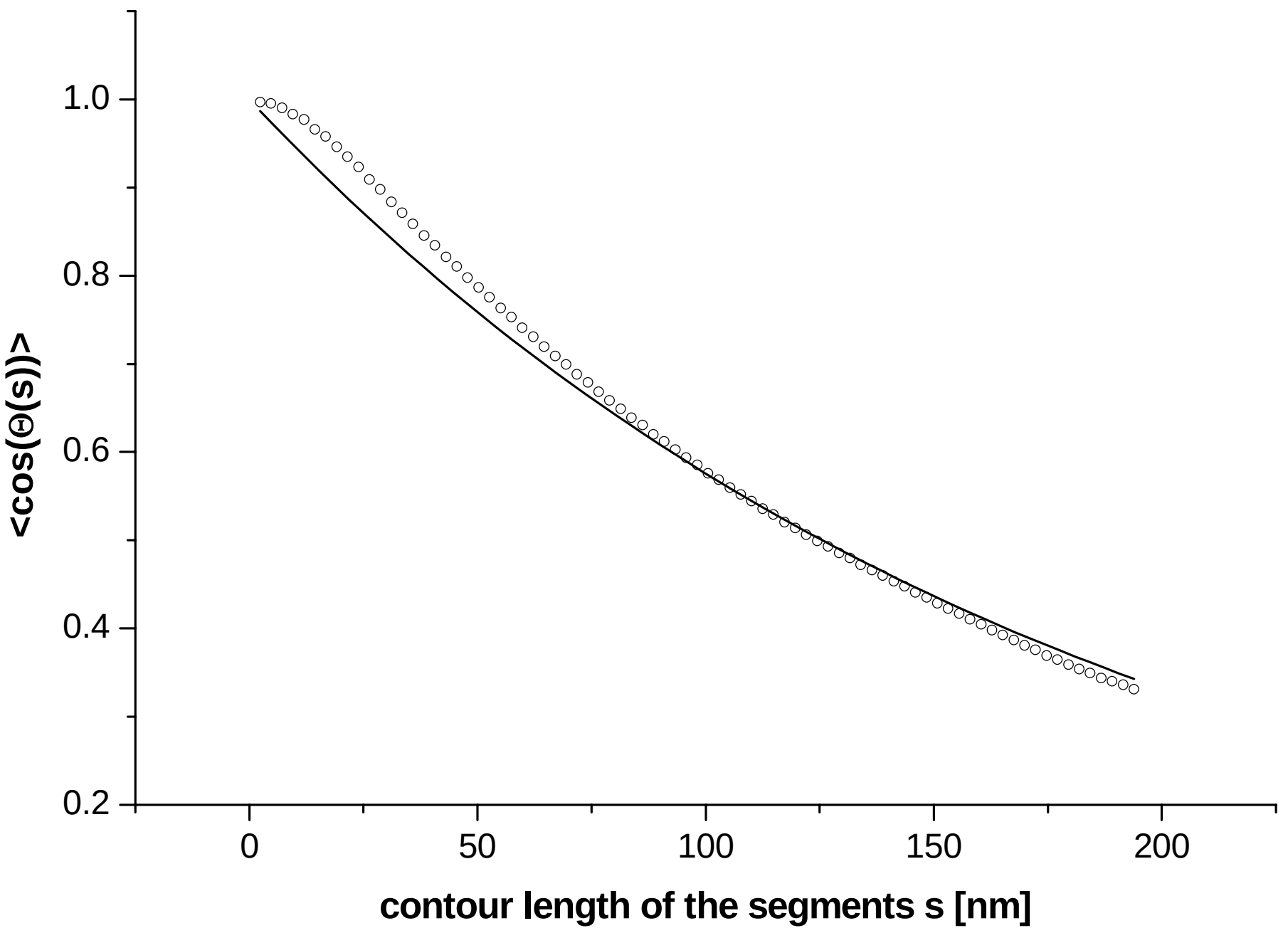

Fig. 6

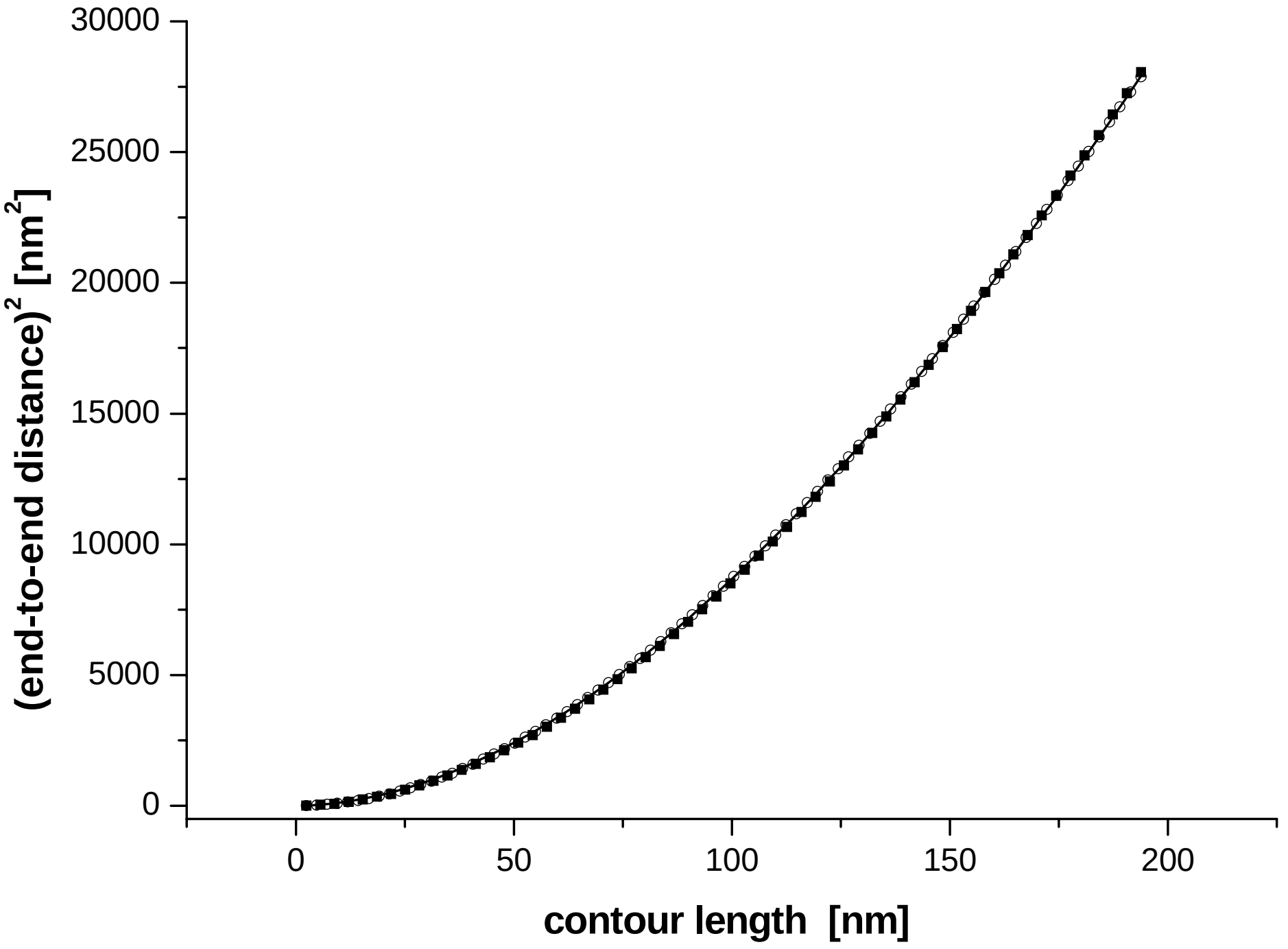

Fig. 7

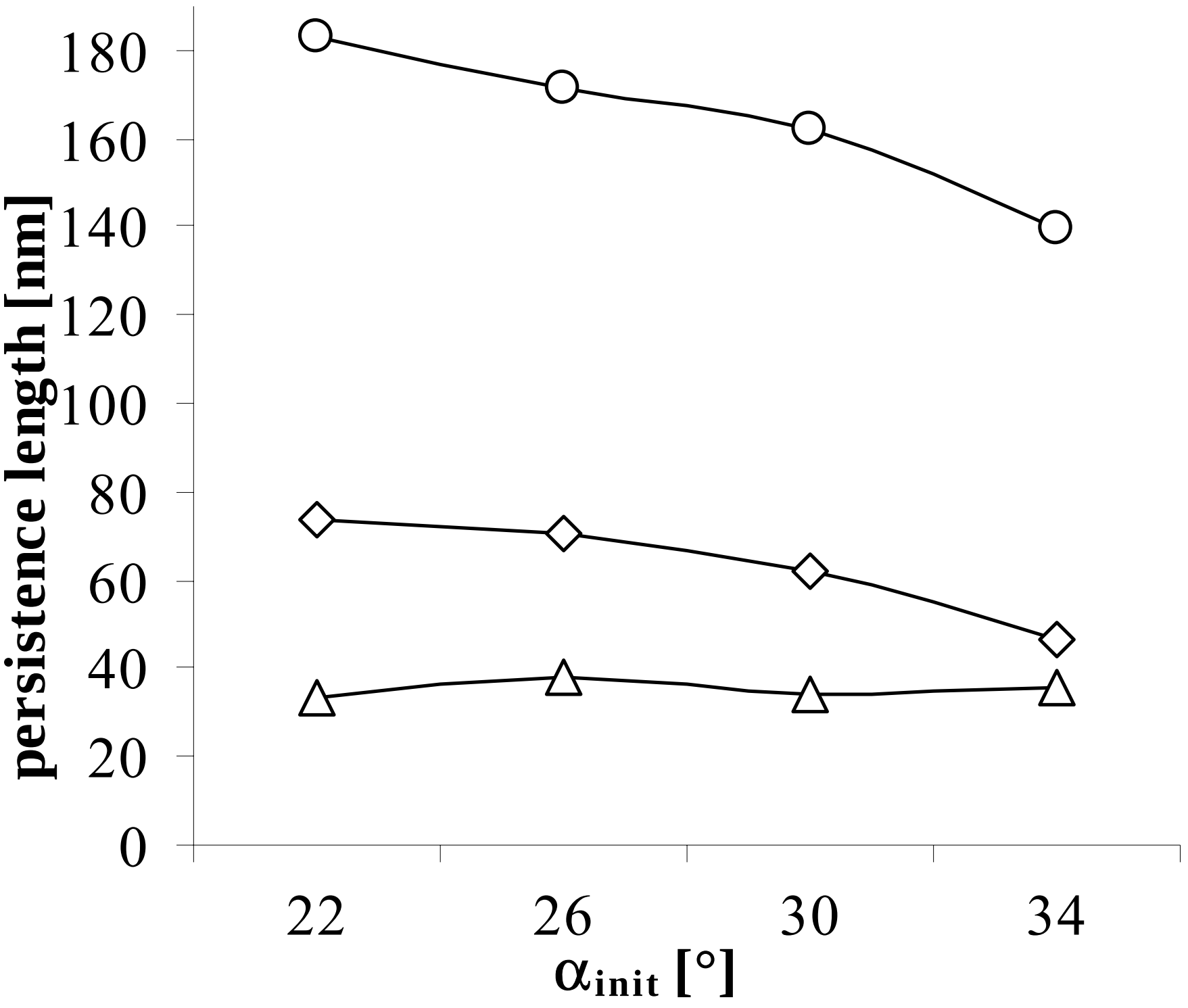

Fig. 8

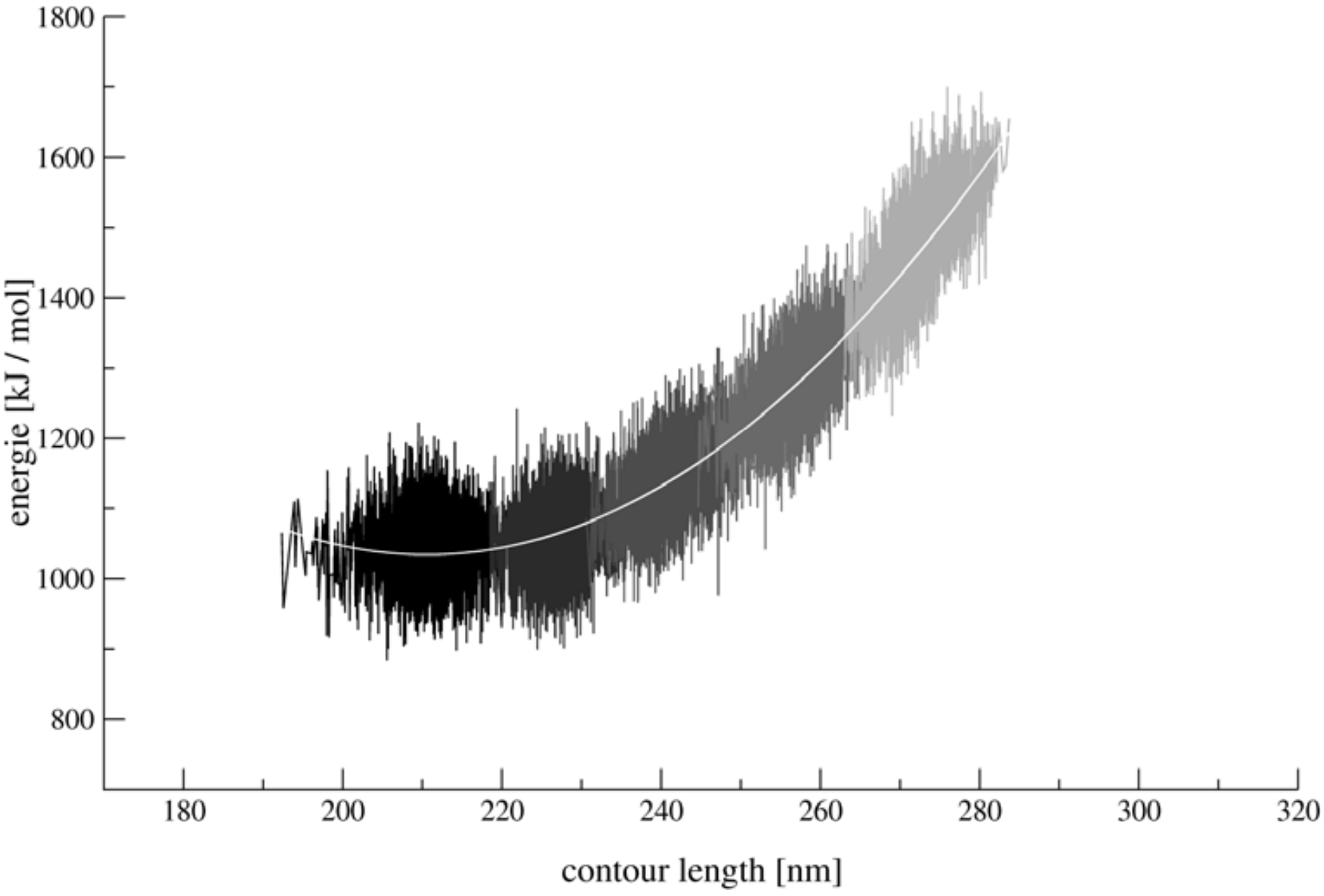

Fig. 9.1

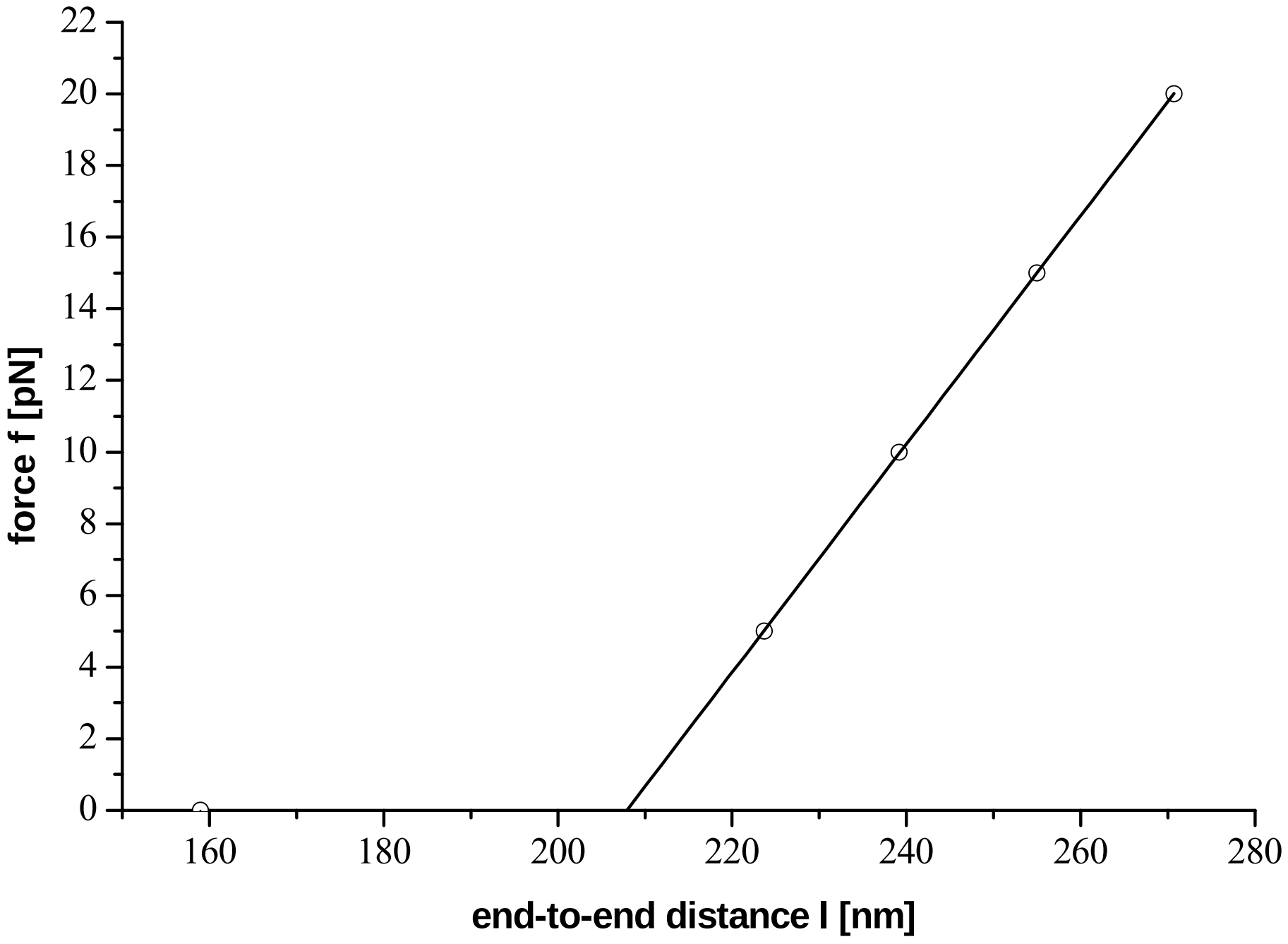

Fig. 9.2

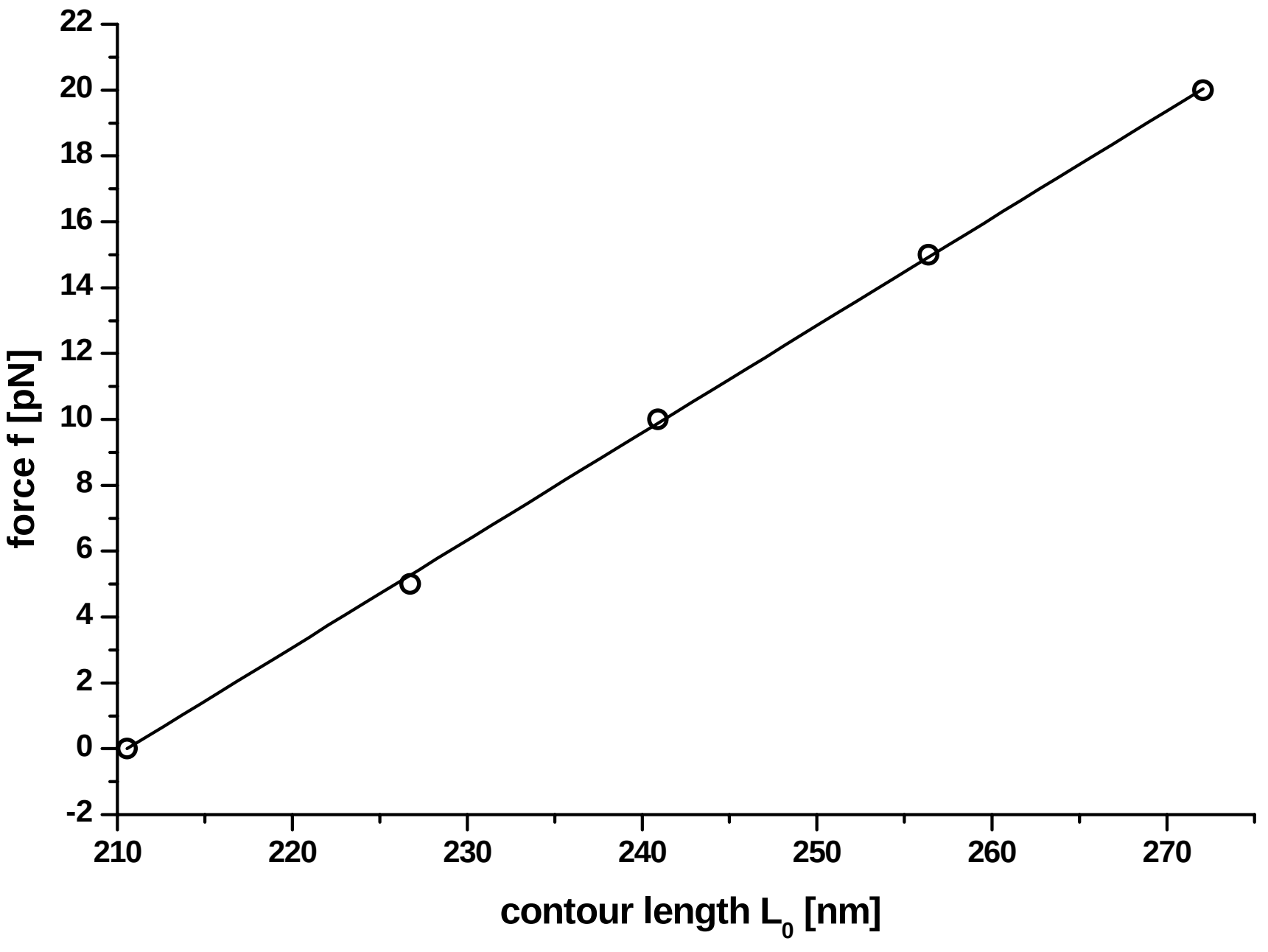

Fig. 9.3

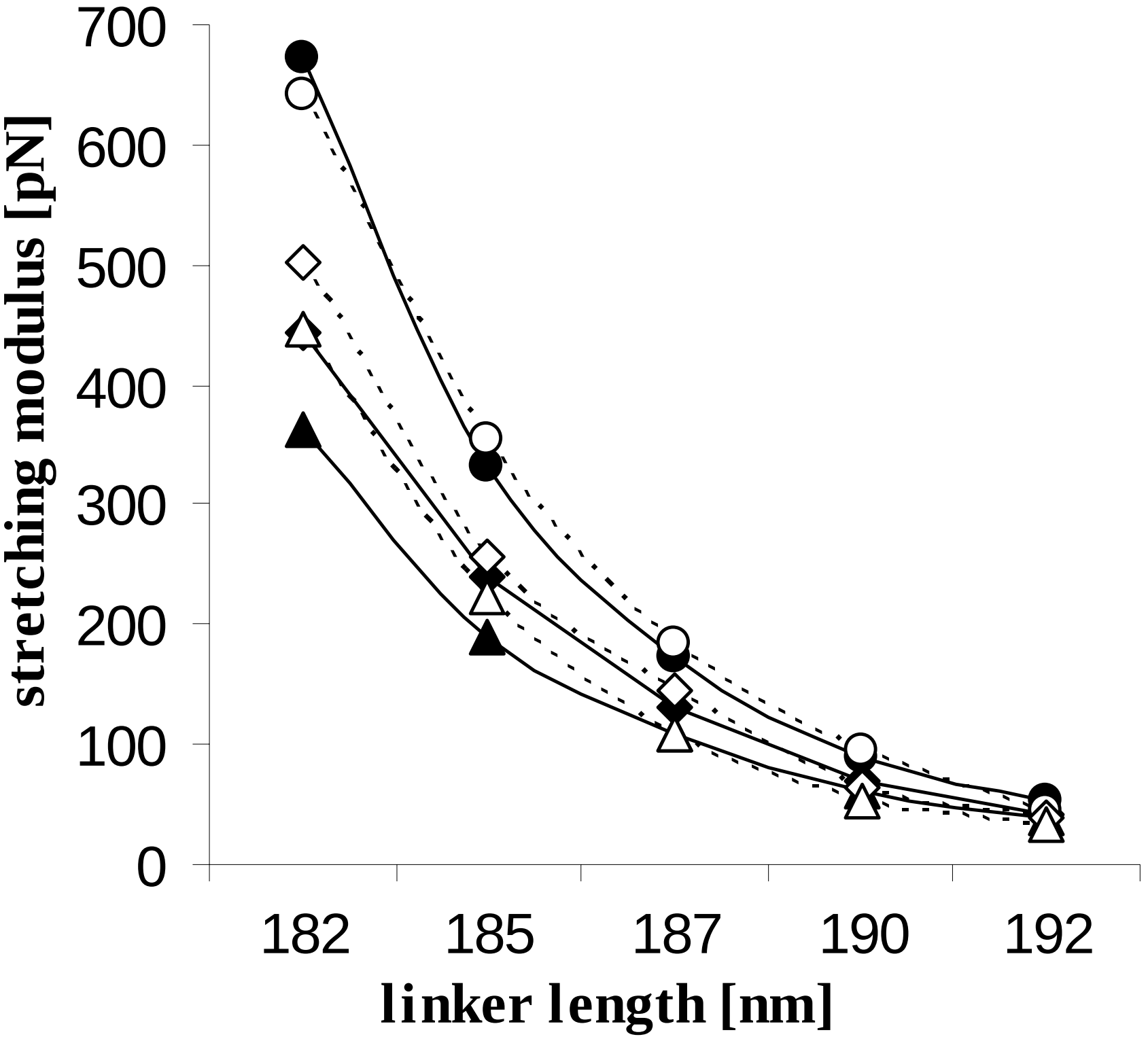

Fig. 10.1

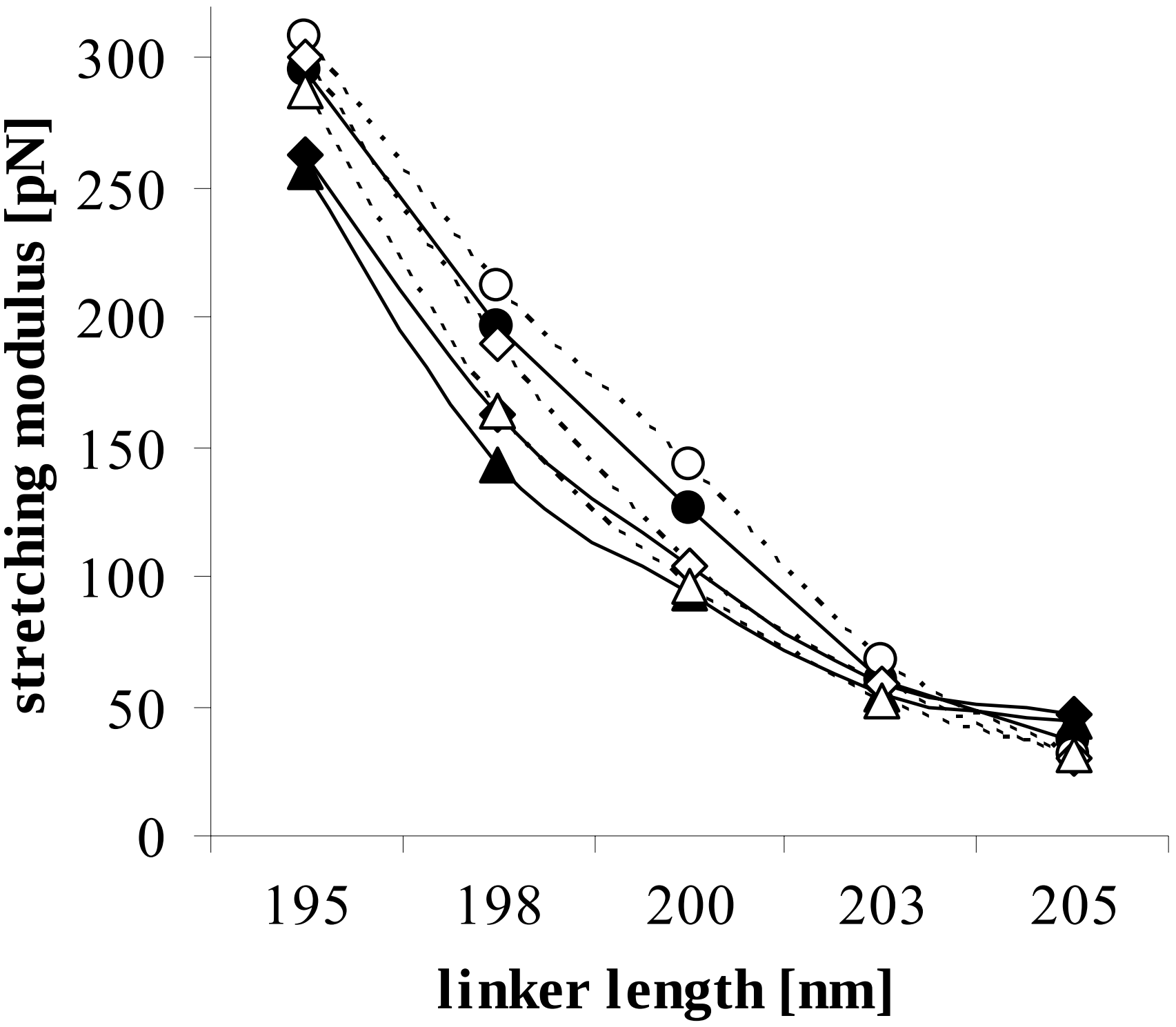

Fig. 10.2

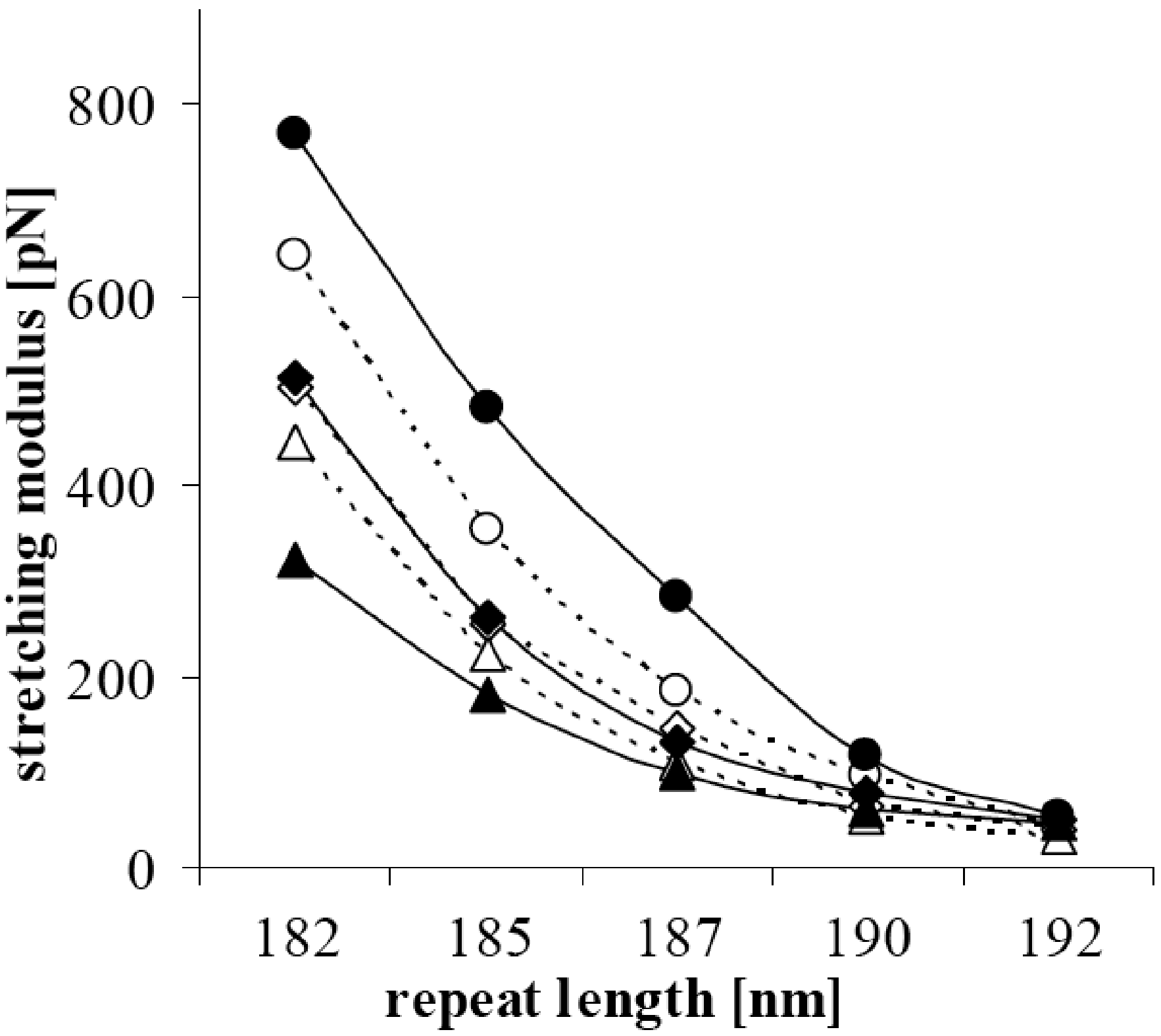

Fig. 11.1

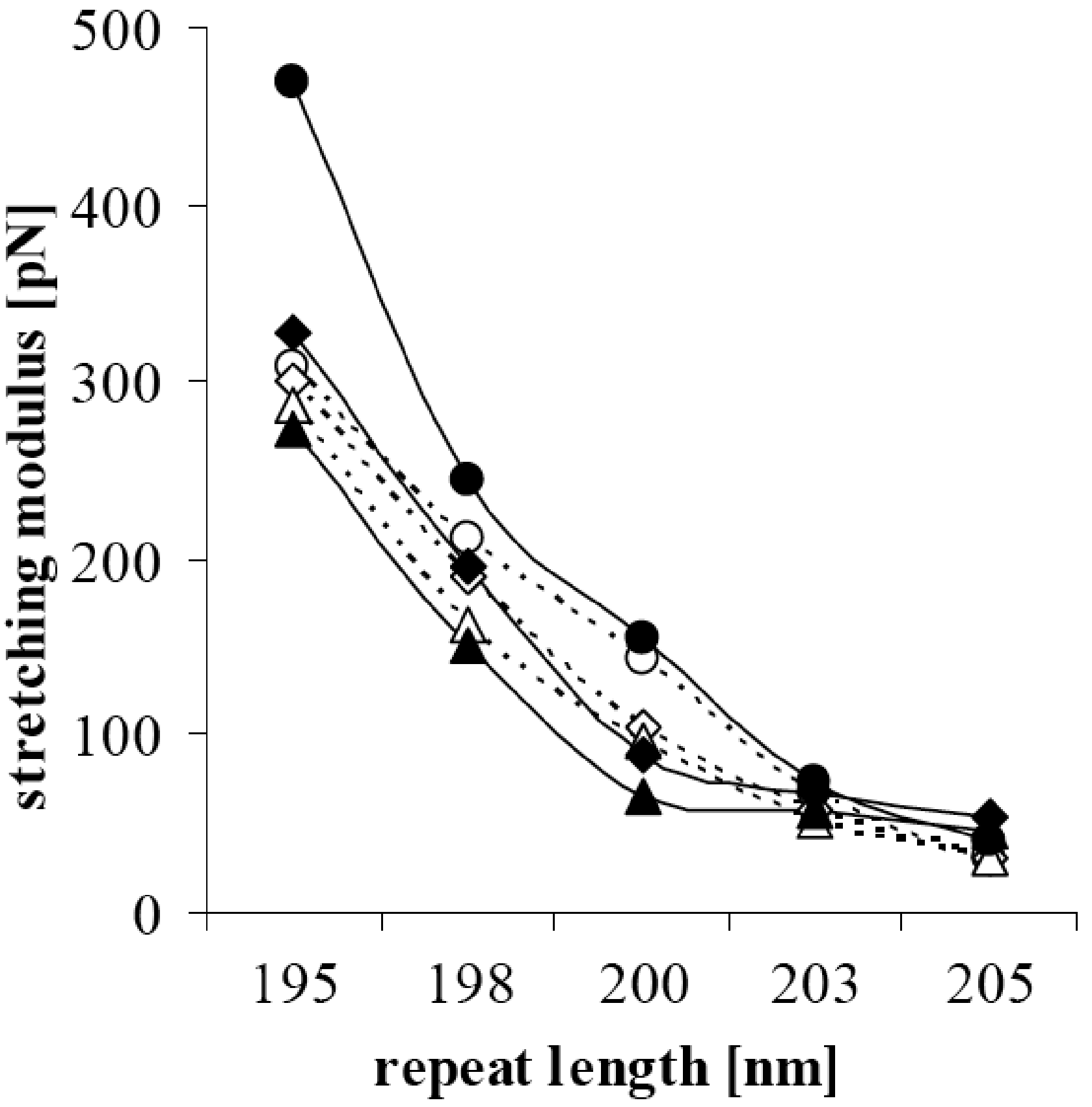

Fig. 11.2

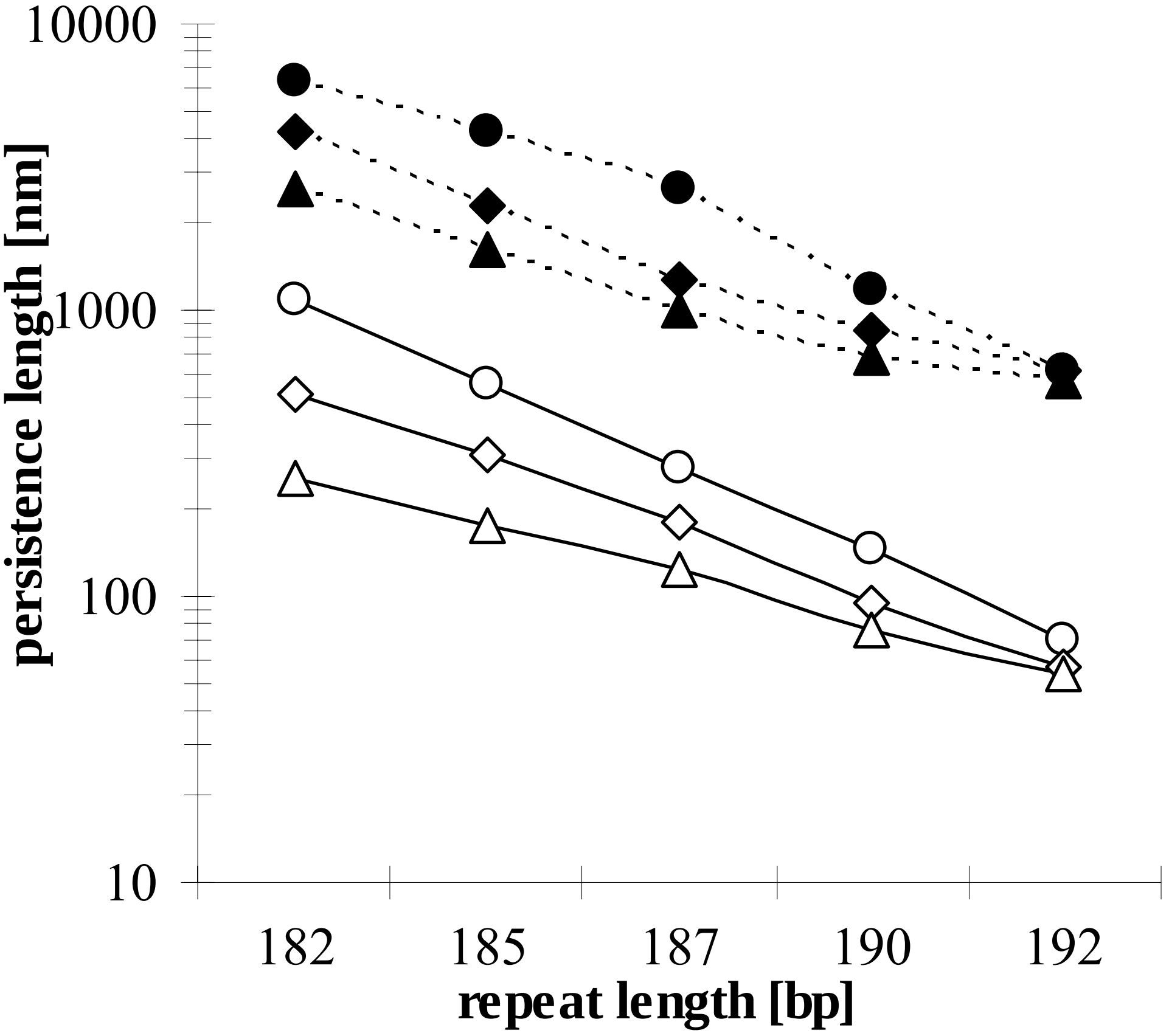

Fig. 12.1

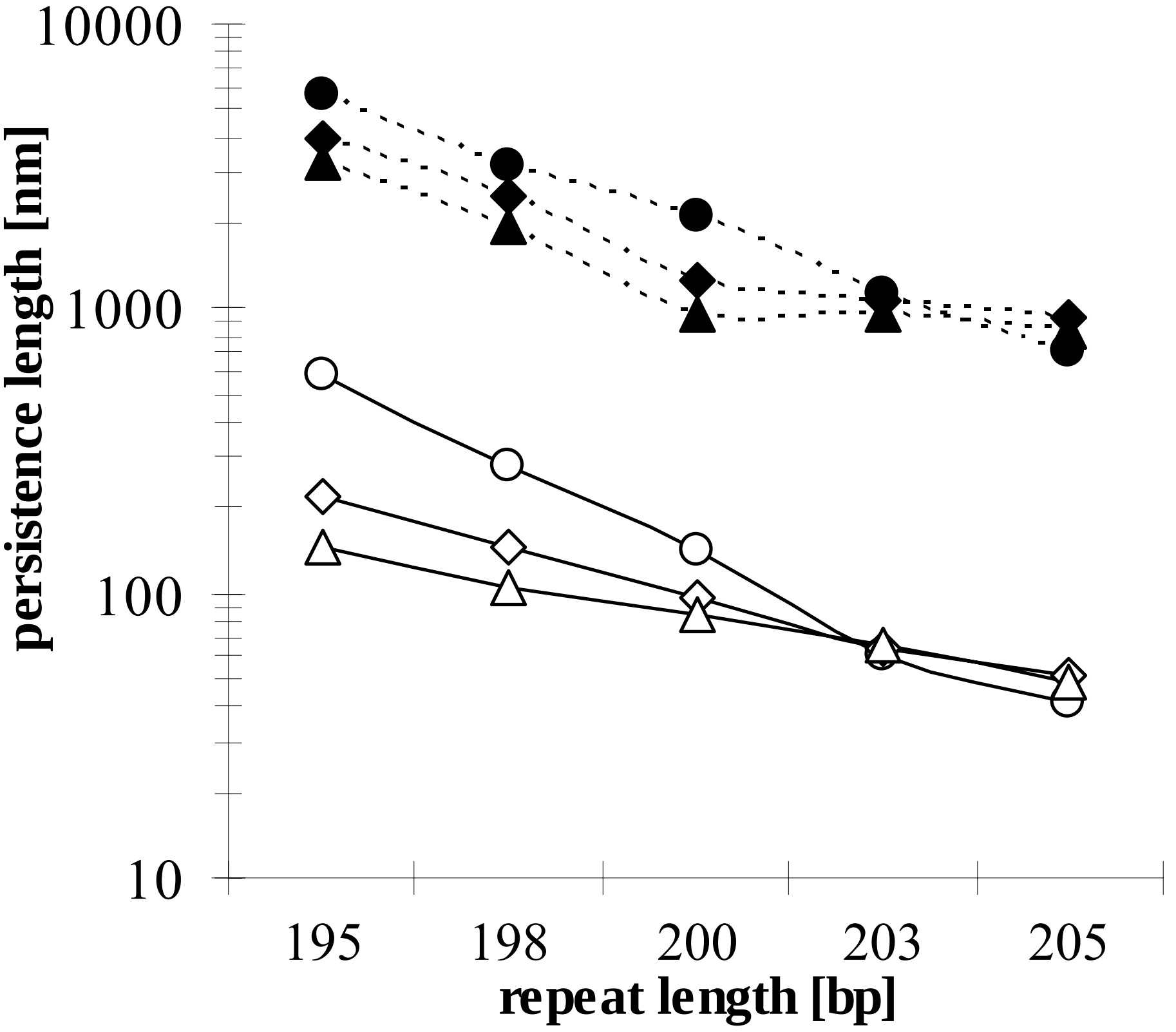

Fig. 12.2

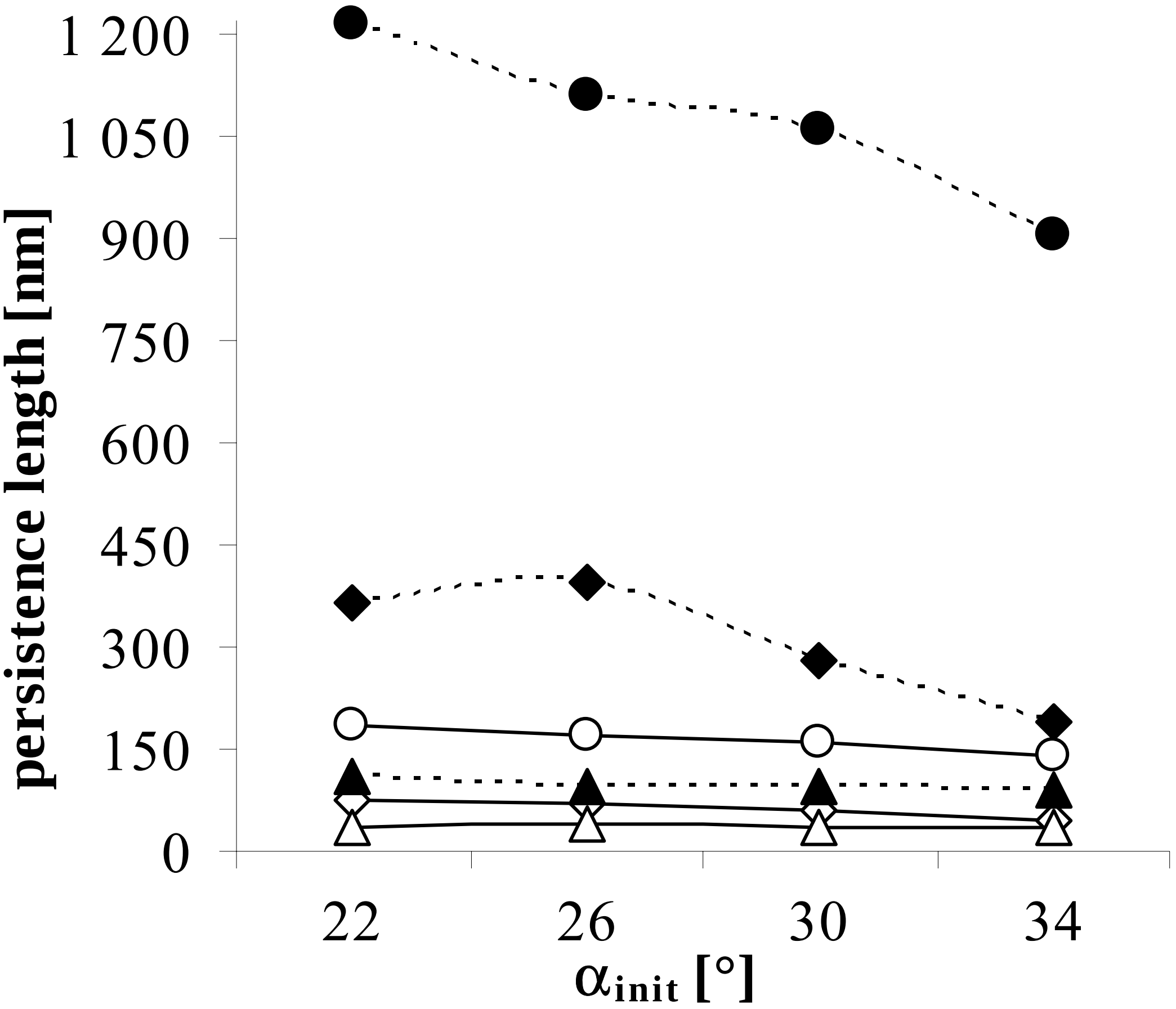

Fig. 12.3

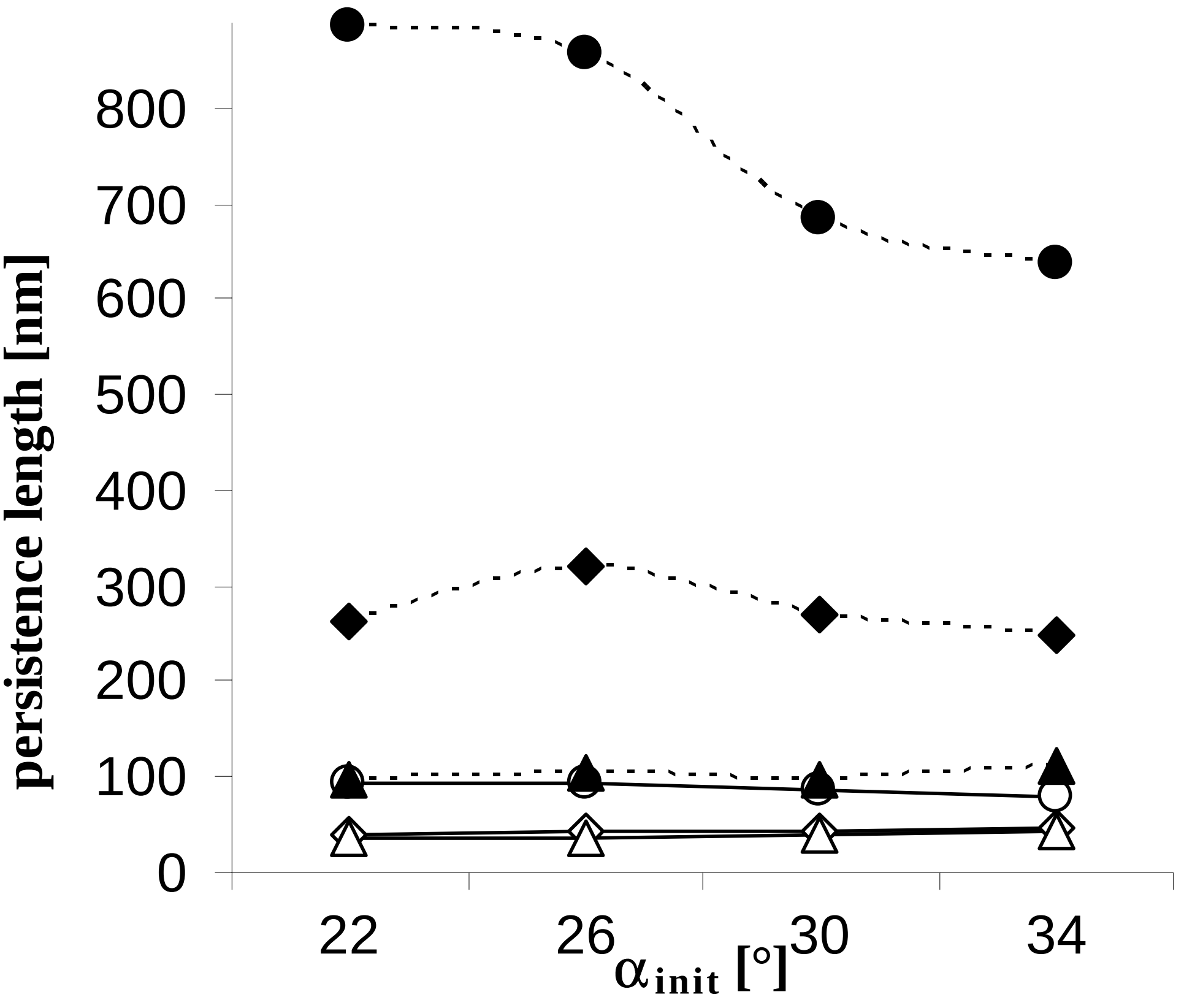

Fig. 12.4

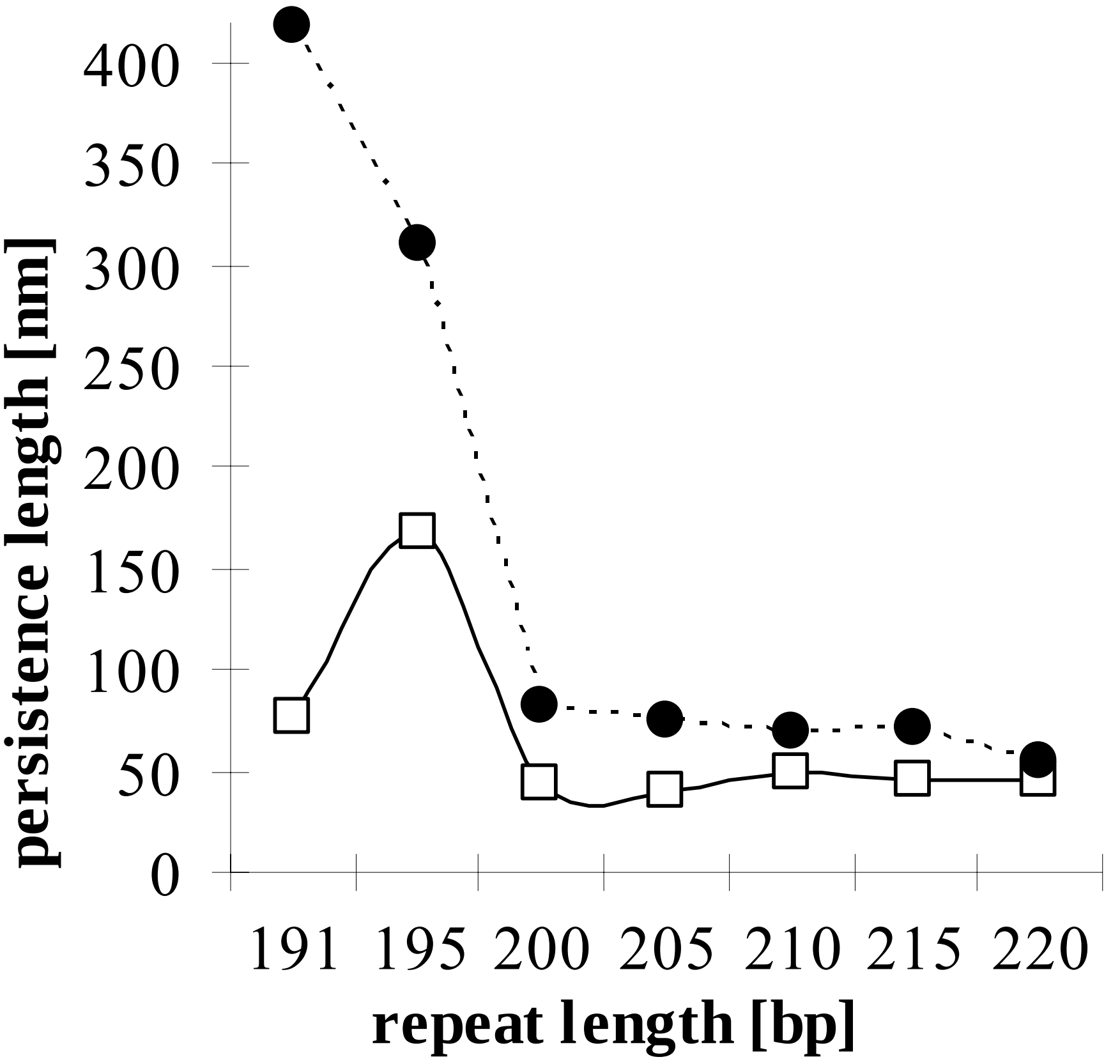

Fig. 13.1

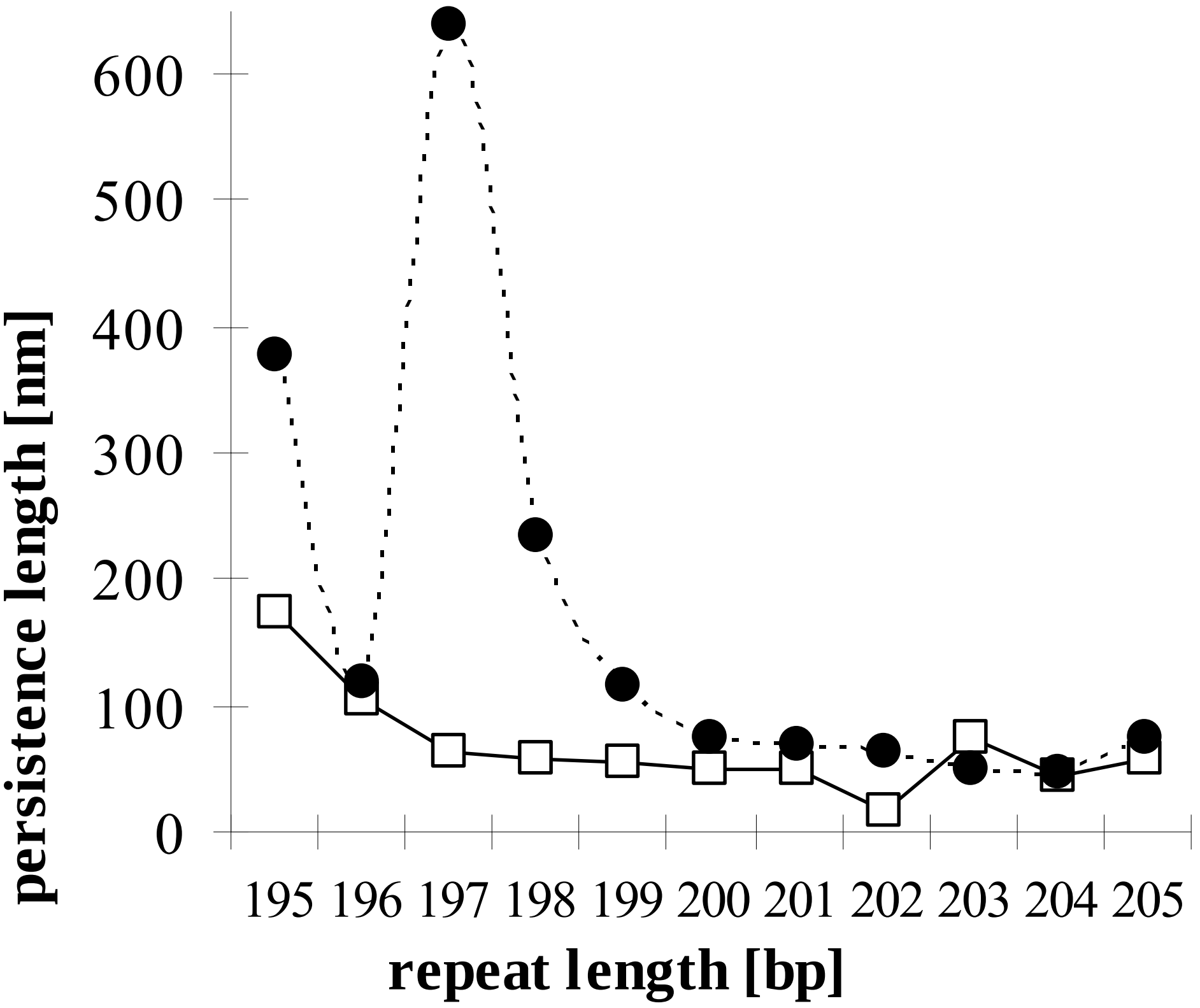
600
500
400
300
200
100
0
persistence length [nm]
195 196 197 198 199 200 201 202 203 204 205
repeat length [bp]

Fig. 13.2